\documentclass[11pt,final,a4paper]{article}
\usepackage{mathpazo}
\usepackage{amsmath,amsthm,amsfonts,amssymb,xcolor}
\usepackage[margin=1in]{geometry}

\usepackage{soul}
\usepackage[multiple]{footmisc}
\usepackage{setspace}
\usepackage{natbib}
\usepackage{tikz}
\usepackage[colorlinks,linkcolor=links,citecolor=cites,urlcolor=MyDarkBlue]{hyperref}
\usepackage{bbm}
\usepackage[justification=centering]{caption}
\usepackage[justification=centering]{subfig}
\usepackage[section]{placeins}
\usepackage{parskip}
\usepackage{nicefrac}

\definecolor{MyDarkBlue}{rgb}{0,0.08,0.45}
\definecolor{cites}{HTML}{324b13}
\definecolor{links}{HTML}{1a663b}
\definecolor{MyLightMagenta}{cmyk}{0.1,0.8,0,0.1}
\newtheorem*{assumptionf}{Assumption \objective}
\newtheorem*{assumptiong}{Assumption \vectorconstraint}
\newtheorem*{assumptioncs}{Assumption S}

\newtheorem{theorem}{Theorem}
\newtheorem{lemma}{Lemma} 
\newtheorem{prop}{Proposition}
 
\newtheorem{example}{Example}
\newtheorem{definition}{Definition} 
\newtheorem{cor}{Corollary}
\newtheorem{assumption}{Assumption}
\newtheorem{observation}{Observation}
\newtheorem{remark}{Remark}
\newcommand{\belief}{\ensuremath{\mu}}
\newcommand{\prior}{\ensuremath{\belief_0}}
\newcommand{\state}{\ensuremath{\omega}}
\newcommand{\stateb}{\ensuremath{\state^\prime}}
\newcommand{\States}{\ensuremath{\Omega}}
\newcommand{\type}{\ensuremath{\theta}}
\newcommand{\typeb}{\ensuremath{\type^\prime}}
\newcommand{\Types}{\ensuremath{\Theta}}
\newcommand{\Posteriors}{\ensuremath{\Delta(\States)}}
\newcommand{\posterior}{\ensuremath{\belief}}

\newcommand{\allocation}{\ensuremath{a}}
\newcommand{\allocationb}{\ensuremath{\allocation^\prime}}
\newcommand{\transfer}{\ensuremath{x}}
\newcommand{\Allocations}{\ensuremath{A}}
\newcommand{\mechanism}{\ensuremath{\mathbf{M}}}
\newcommand{\inputt}{\ensuremath{M}}
\newcommand{\outputt}{\ensuremath{S}}
\newcommand{\joint}{\ensuremath{\varphi}}
\newcommand{\constraint}{\ensuremath{\mathcal{\Allocations}}}
\newcommand{\measurablem}{\ensuremath{\tilde{U}}}
\newcommand{\measurablet}{\ensuremath{\tilde{\States}}}
\newcommand{\jointp}{\ensuremath{\mathbb{P}}}
\newcommand{\action}{\ensuremath{a}}

\newcommand{\experiment}{\ensuremath{\pi}}
\newcommand{\experimentb}{\ensuremath{\experiment^\prime}}

\newcommand{\genericf}{\ensuremath{h}}

\newcommand{\reals}{\ensuremath{\mathbb{R}}}
\newcommand{\naturals}{\ensuremath{\mathbb{N}}}
\newcommand{\conjugate}{\ensuremath{^*}}
\newcommand{\dual}{\ensuremath{t}}
\newcommand{\variable}{\ensuremath{x}}
\newcommand{\weight}{\ensuremath{\lambda}}
\newcommand{\closure}{\ensuremath{\mathrm{cl}}}
\newcommand{\conv}{\ensuremath{\mathrm{conv}}}
\newcommand{\cav}{\ensuremath{\mathrm{cav}}}
\newcommand{\clcav}{\ensuremath{\overline{\cav}}}
\newcommand{\vectorconstraint}{\ensuremath{\overline{\const}}}
\newcommand{\const}{\ensuremath{g}}
\newcommand{\supp}{\mathop{\mathrm{supp}}}

\newcommand{\objective}{\ensuremath{f}}
\newcommand{\modifiedobjective}{\ensuremath{\objective^{\vectorconstraint}}}
\newcommand{\equality}{\ensuremath{E}}
\newcommand{\inequality}{\ensuremath{I}}
\newcommand{\price}{\ensuremath{p}}
\newcommand{\side}{\ensuremath{\gamma}}
\newcommand{\sideindex}{\ensuremath{\tilde{\side}}}
\newcommand{\variableindex}{\ensuremath{\tilde{\variable}}}

\newcommand{\dom}{\ensuremath{\mathrm{dom}}}

\newcommand{\interior}{\ensuremath{\mathrm{int}}}

\newcommand{\hypograph}{\ensuremath{\mathrm{hypo}}}
\newcommand{\graph}{\ensuremath{\mathrm{graph}}}
\newcommand{\level}{\ensuremath{y}}
\newcommand{\Value}{\ensuremath{V}}
\newcommand{\feasible}{\ensuremath{\mathcal{F}}}
\newcommand{\set}{\ensuremath{S}}
\newcommand{\constraintset}{\ensuremath{C}}
\newcommand{\binding}{\ensuremath{B}}

\newcommand{\bsplit}{\ensuremath{\tau}}
\newcommand{\bsplitopt}{\ensuremath{\bsplit^\star}}
\newcommand{\bsplitb}{\ensuremath{\bsplit^{\star\star}}}
\newcommand{\vectorconstraintb}{\ensuremath{\vectorconstraint^\prime}}
\newcommand{\sideb}{\ensuremath{\side^\prime}}
\newcommand{\objectiveopt}{\ensuremath{\objective^\star}}
\newcommand{\vectorconstraintopt}{\ensuremath{\vectorconstraint^\star}}
\newcommand{\supportset}{\ensuremath{A}}
\newcommand{\convexset}{\ensuremath{B}}

\newcommand{\beliefopt}{\ensuremath{\belief^\star}}

\newcommand{\beliefoptb}{\ensuremath{\belief^{\star\star}}}

\newcommand{\jointcav}{\ensuremath{J}}
\newcommand{\canonical}{\ensuremath{\mathrm{e}}}
\newcommand{\canonicalobj}{\ensuremath{\canonical_\objective}}
\newcommand{\canonicalconst}{\ensuremath{\canonical_i}}
\newcommand{\feasibleset}{\ensuremath{F}}
\newcommand{\lagrangian}{\ensuremath{L}}
\newcommand{\dimension}{\ensuremath{D}}
\newcommand{\jointf}{\ensuremath{j}}

\newcommand{\bestresponse}{\ensuremath{BR}}
\newcommand{\beliefmatrix}{\ensuremath{B}}
\newcommand{\support}{\ensuremath{\mathrm{supp}}}

\newcommand{\OPT}{CID}

\usepackage{chngcntr}
\numberwithin{equation}{section}
\numberwithin{theorem}{section}
\numberwithin{definition}{section}
\numberwithin{example}{section}
\numberwithin{cor}{section}
\numberwithin{prop}{section}
\numberwithin{lemma}{section}
\usepackage{tikz}
\usepackage{pgfplots}
\pgfplotsset{compat=1.16}
\usepgfplotslibrary{fillbetween}
\usepgfplotslibrary{patchplots}
\usetikzlibrary{positioning,chains,fit,shapes,calc,arrows,trees,decorations.pathmorphing,decorations.markings,hobby,calc,snakes,shapes.misc,patterns,backgrounds,quotes}
\usetikzlibrary{arrows}
\usetikzlibrary{trees,calc,snakes,fit,shapes,positioning}
\usetikzlibrary{decorations.pathmorphing}
\usetikzlibrary{decorations.markings}
\usetikzlibrary{decorations.shapes}
\tikzset{cross/.style={cross out, draw=black, minimum size=2*(#1-\pgflinewidth), inner sep=0pt, outer sep=0pt},
cross/.default={1pt}}

 \tikzset{ every node/.style={inner sep=0pt,minimum size=1mm},
  nsnode/.style={draw,circle,black},
    wsnode/.style={draw,circle,white},
  nnnode/.style={draw,circle,black,fill=black},
  asnode/.style={draw,circle,blue,fill=blue},
    asnode2/.style={draw,circle,blue},
      rsnode/.style={draw,circle,red,fill=red},
    rsnode2/.style={draw,circle,red},
  bsnode/.style={draw,circle,mygreen,fill=mygreen},
  csnode/.style={draw,circle,red,fill=red, minimum size=2mm},
  every fit/.style={inner sep=-1.5pt,text width=1cm}  }
 \usepackage{tcolorbox}
 \newtcolorbox{mybox}{width=0.45\textwidth,enhanced,colback=red!5!white,colframe=red!75!black}
 \tikzset{
  treenode/.style = {align=center, inner sep=0pt, text centered,
    font=\sffamily},
  arn_n/.style = {treenode, circle, black, font=\sffamily\bfseries, draw=black,
    fill=white, text width=1.5em},
  arn_r/.style = {treenode, circle, newblue,font=\sffamily\bfseries,draw=newblue, 
    text width=1.5em},
  arn_x/.style = {treenode, circle, orange,draw=orange,font=\sffamily\bfseries,
   text width=2em}
}
 \tikzstyle{block} = [draw=orange, fill=bg, rectangle,rounded corners,
    minimum height=4em, minimum width=6em,align=center,text width=3 cm]
\usepackage{tikzsymbols}
\usepackage{xr}
\newcommand\fnsep{\textsuperscript{,}}
\title{Constrained Information Design\thanks{The latest version of this paper can be found \href{https://www.dropbox.com/s/as70bw8c5flneoe/cidt-public.pdf?dl=0}{here}. We thank the Editor, Jos\'e Correa, an Associate Editor, and four anonymous referees for feedback that has greatly improved the paper. William Grimme provided excellent research assistance. This research is supported by grants from the National Science Foundation (Doval: SES-2131706; Skreta: SES-1851729). Vasiliki Skreta is grateful for generous financial support through the ERC consolidator grant 682417 ``Frontiers in design.''}}
\author{Laura Doval\thanks{Columbia Business School and CEPR. E-mail: \href{mailto:laura.doval@columbia.edu}{\texttt{laura.doval@columbia.edu}}} \and Vasiliki Skreta\thanks{University of Texas at Austin, University College London, and CEPR. E-mail: \href{mailto:vskreta@gmail.com}{\texttt{vskreta@gmail.com}}.}}

\begin{document}
\pagenumbering{gobble}
\maketitle
\begin{abstract}
We provide tools to analyze information design problems subject to constraints. We do so by extending the insight in \cite{le2019persuasion} to the case of multiple inequality and equality constraints. Namely, that an information design problem subject to constraints can be represented as an unconstrained information design problem with a \emph{additional} states, one for each constraint. Thus, without loss of generality, optimal solutions induce as many posteriors as the number of states \emph{and} constraints. We provide results that refine this upper bound. Furthermore, we provide conditions under which there is no duality gap in constrained information design, thus validating a Lagrangian approach. We illustrate our results with applications to mechanism design with limited commitment (\citealp{doval2020mechanism}) and persuasion of a privately informed receiver (\citealp{kolotilin2017persuasion}).

\end{abstract}
\normalsize
\newpage
\clearpage
\pagenumbering{arabic}
\section{Introduction}\label{sec:intro}
%
We provide tools to solve constrained information design problems. These problems are becoming common: Since \cite{kamenica2011bayesian} seminal paper on Bayesian persuasion, the literature on information design has grown steadily. A bulk of new work analyzes \emph{constrained} information design problems, which can be classified in three groups:
\begin{enumerate}
\item The information designer faces constraints additional to the Bayes' plausibility constraint in \cite{kamenica2011bayesian}, like participation in \cite{rosar2017test}, moral hazard in \cite{boleslavsky2018bayesian}, and capacity constraints on information transmission in \cite{le2019persuasion}.

\item  The designer is designing an information structure that is part of a mechanism that satisfies incentive and participation constraints, like in mechanism design with aftermarkets (e.g., \citealp{calzolari2006monopoly,dworczak2020mechanism}), and in mechanism design with limited commitment (e.g., \citealp{doval2020mechanism}).
%
%
\item Mechanism design problems that do not involve information design and still can be solved using information design tools, like in  \cite{georgiadis2020optimal} and \cite{dworczak2019redistribution}.
\end{enumerate} 
A natural approach to tackle these constrained information design problems is to set up a Lagrangian to incorporate the constraints into the objective function, except for the Bayes' plausibility constraint. If each constraint can be written as the expectation over posteriors of some function, then the Lagrangian itself can be written as an expectation over posteriors of some function given the Lagrange multiplier. If there are $N$ possible states of the world, one may be tempted to apply a corollary of Carath\'eodory's theorem (Corollary 17.1.5 in \citealp{rockafellar2015convex}) \label{page-ct-cor-intro} and conclude from this that the optimal information policy uses at most $N$ posteriors. After all, the solution to the problem would correspond to the concavification of the function whose expectation over posteriors determines the Lagrangian. 

However, as we illustrate next, this approach may fail to deliver the correct solution to the constrained information design problem:
\begin{example}[Naive Lagrangian approach]\label{example:lobbying} Consider the following constrained information design problem. There are two equally likely states of the world $\state\in\{0,1\}\equiv\States$. Letting $\belief\in[0,1]$ denote the probability that the state is $\state=1$, the objective function is given by:
%
\begin{align}\label{eq:lobbying-objective}
\objective(\belief)=\left\{\begin{array}{ll}0 & \text{ if }\belief<\frac{1}{3}\\
m&\text{ if } \belief\in[\frac{1}{3},\frac{2}{3})\\
1&\text{otherwise}\end{array}\right.,
\end{align}
where $m$ is a parameter $m\in[0,0.5)$.  Denoting by $\Delta(\Posteriors)$ the set of posterior distributions, we intend to find a posterior distribution \bsplit\ with mean $\prior=\nicefrac{1}{2}$ that maximizes the expectation of \objective\ subject to a constraint defined by the function \const, where 
\begin{align}\label{eq:lobbying-constraint}
\const(\belief)=\mathbbm{1}\left[\belief\in[1/3,2/3]\right].
\end{align}
The constrained information design problem is then:
\begin{align}\tag{\OPT$_L$}\label{eq:lobbying-opt}
\max_{\bsplit\in\Delta(\Posteriors)}&\mathbbm{E}_\bsplit\left[\objective(\belief)\right]\\
\text{ s.t. }&\left\{\begin{array}{l}\mathbb{E}_{\bsplit}\left[\belief\right]=\prior \\
\mathbbm{E}_\bsplit\left[\const(\belief)\right]\geq\side\end{array}\right..\nonumber
\end{align}
That is, we choose a posterior distribution \bsplit\ with mean \prior\ to maximize the expectation of the objective subject to the constraint that in expectation the value of \const\ is at least \side, where $\side\in(3/4,1)$. Letting $\dual\geq0$ denote the multiplier on the inequality constraint, we define the Lagrangian objective:
\begin{align}
(\objective+\dual\const)(\belief)=\left\{\begin{array}{ll}0 & \text{ if }\belief<\frac{1}{3}\\
m+\dual&\text{ if } \belief\in[\frac{1}{3},\frac{2}{3})\\
1+\dual&\text{ if }\belief=\frac{2}{3}\\
1&\text{otherwise}\end{array}\right..
\end{align}
It is then natural to set up the following problem:
\begin{align}\label{eq:lobbying-lagrangian}
\min_{t\geq0}\max_{\bsplit\in\Delta\left(\Posteriors\right)}&\mathbb{E}_\bsplit\left[\objective(\belief)+\dual\const(\belief)\right]\\
\text{ s.t. }&\mathbb{E}_\bsplit\left[\belief\right]=\prior.\nonumber
\end{align}
Appealing to the results in \cite{aumann1995repeated,kamenica2011bayesian}, one may recognize that the solution to the inner-maximization in \autoref{eq:lobbying-lagrangian} corresponds to the \emph{concave hull} of $\objective+\dual\const$ at \prior\ (see \autoref{definition:concave-hull}) and rely on standard results to reduce the search for an optimal posterior distribution to those distributions with at most binary support. Visual inspection of the Lagrangian objective and its concavification in \autoref{fig:lobbying-lagrangian} reveal that this would not be an issue as long as the multiplier \dual\ is different from $\nicefrac{1}{2}$. Indeed, for the purposes of concavifying the Lagrangian at the prior \prior\ distributions with at most binary support suffice. For instance, when $\dual<\nicefrac{1}{2}$, the optimal distribution splits \prior\ between $0$ and $\nicefrac{2}{3}$, which leads to an infeasible solution (see \autoref{fig:lobbying-low-dual}). Instead, when $\dual>\nicefrac{1}{2}$, the optimal distribution splits \prior\ between $\nicefrac{1}{3}$ and $\nicefrac{2}{3}$, which satisfies the constraint with slack, contradicting that $\dual>0$. 

When $\dual=\nicefrac{1}{2}$, multiple distributions over posteriors  attain the concavification of the Lagrangian objective at the prior. Consistent with Corollary 17.1.5 in \cite{rockafellar2015convex}, there are binary support distributions that attain the concavification. For instance, we can split the prior between $\nicefrac{1}{3}$ and $\nicefrac{2}{3}$ or we could split the prior between $0$ and $\nicefrac{2}{3}$. The former satisfies the constraints, but delivers the no disclosure payoff to the objective function. Instead, the latter, which delivers the optimal payoff for the objective \objective, does not satisfy the constraints. However, there is a third distribution supported on three points, $\{0,\nicefrac{1}{3},\nicefrac{2}{3}\}$, that attains the concavification of the Lagrangian objective, while at the same time satisfying the constraints: it balances the desire of \objective\ to (partially) reveal information with the satisfaction of the inequality constraint. Note that all three splittings of the prior deliver the same value for the Lagrangian objective, but they are not all feasible nor optimal in the constrained information design problem, \ref{eq:lobbying-opt}. 

For this example, the solution to the program in \autoref{eq:lobbying-lagrangian} is given by $\dual\conjugate=0.5$ and $\{(\tau^\star(\beliefopt_m),\beliefopt_m)\}_{m=1}^3=\{(1-\side,0),(2\side-\nicefrac{3}{2},\nicefrac{1}{3}),(\nicefrac{3}{2}-\side,\nicefrac{2}{3})\}$. 
\begin{figure}[h!]
\centering
\subfloat[Objective function \objective]{\begin{tikzpicture}[scale=0.8]\begin{axis}[axis lines=middle,
            xmin=-0.05,xmax=1,
        ymin=-0.05,ymax=1.5,
        xtick={0.33,0.5,0.66},
                xticklabels={$\frac{1}{3}$,$\prior$,$\frac{2}{3}$},
        ytick={0,0.25,1},
        yticklabels={0,$m$,$1$},
        xlabel=$\belief$,ylabel=$\objective$,
         x label style={at={(axis description cs:1,-0.05)}},
    y label style={at={(axis description cs:-0.05,1)}},
    width=8cm,
        height=8cm
]
\addplot[color=black,very thick]coordinates {(0,0) (1/3,0)};
\addplot[color=black,mark=o] coordinates {(1/3,0)};
\addplot[color=black,mark=*] coordinates {(1/3,0.25)};
\addplot[color=black,very thick] coordinates {(1/3,0.25) (2/3,0.25)};
\addplot[color=black,mark=o] coordinates {(2/3,0.25)};
\addplot[color=black,mark=*] coordinates {(2/3,1)};
\addplot[color=black,very thick] coordinates {(0.66,1) (1,1)};
\addplot[color=red,dashed,very thick] coordinates {(0,0) (2/3,1)};
\addplot[color=red,dashed,very thick] coordinates { (2/3,1) (1,1)};
\end{axis}
\end{tikzpicture}\label{fig:lobbying-objective}}
\subfloat[Lagrangian objective, \dual=\dual\conjugate]{\begin{tikzpicture}[scale=0.8]\begin{axis}[axis lines=middle,
            xmin=-0.05,xmax=1,
        ymin=-0.05,ymax=1.5,
        xtick={0.33,0.5,0.66},
                xticklabels={$\frac{1}{3}$,$\prior$,$\frac{2}{3}$},
        ytick={0,0.75,1},
        yticklabels={0,$m+\dual$,$1$},
        xlabel=$\belief$,ylabel=$\objective+\dual\conjugate\const$,
         x label style={at={(axis description cs:1,-0.05)}},
    y label style={at={(axis description cs:-0.2,1)}},
    width=8cm,
        height=8cm
]
\addplot[color=blue,very thick]coordinates {(0,0) (1/3,0)};
\addplot[color=blue,mark=o] coordinates {(1/3,0)};
\addplot[color=blue,mark=*] coordinates {(1/3,0.75)};
\addplot[color=blue,very thick] coordinates {(1/3,0.75) (2/3,0.75)};
\addplot[color=blue,mark=*] coordinates {(2/3,1.5)};
\addplot[color=blue,mark=o] coordinates {(2/3,1)};
\addplot[color=blue,very thick] coordinates {(2/3,1) (1,1)};
\addplot[color=red,dashed,thick] coordinates {(0,0) (2/3,1.5)};
\addplot[color=red,dashed,thick] coordinates {(2/3,1.5) (1,1)};
\end{axis}
\end{tikzpicture}\label{fig:lobbying-opt-dual}}

\subfloat[Lagrangian objective, $\dual=0.25<\dual\conjugate$]{\begin{tikzpicture}[scale=0.8]
\begin{axis}[axis lines=middle,
            xmin=-0.05,xmax=1,
        ymin=-0.05,ymax=1.5,
        xtick={0.33,0.5,0.66},
                xticklabels={$\frac{1}{3}$,$\prior$,$\frac{2}{3}$},
        ytick={0,0.5,1},
        yticklabels={0,$m+\dual$,$1$},
        xlabel=$\belief$,ylabel=$\objective+\dual\const$,
         x label style={at={(axis description cs:1,-0.05)}},
    y label style={at={(axis description cs:-0.2,1)}},
    width=8cm,
        height=8cm
]
\addplot[color=blue,very thick]coordinates {(0,0) (1/3,0)};
\addplot[color=blue,mark=o] coordinates {(1/3,0)};
\addplot[color=blue,mark=*] coordinates {(1/3,0.55)};
\addplot[color=blue,very thick] coordinates {(1/3,0.55) (2/3,0.55)};
\addplot[color=blue,mark=*] coordinates {(2/3,1.25)};
\addplot[color=blue,mark=o] coordinates {(2/3,1)};
\addplot[color=blue,very thick] coordinates {(2/3,1) (1,1)};
\addplot[color=red,dashed,thick] coordinates {(0,0) (2/3,1.25)};
\addplot[color=red,dashed,thick] coordinates {(2/3,1.25) (1,1)};
\end{axis}
\end{tikzpicture}\label{fig:lobbying-low-dual}}
\subfloat[Lagrangian objective, $\dual=0.8>\dual\conjugate$]{\begin{tikzpicture}[scale=0.8]\begin{axis}[axis lines=middle,
            xmin=-0.05,xmax=1,
        ymin=-0.05,ymax=2,
        xtick={0.33,0.5,0.66},
                xticklabels={$\frac{1}{3}$,$\prior$,$\frac{2}{3}$},
        ytick={0,1.05,1.8},
        yticklabels={0,$m+\dual$,$1+\dual$},
        xlabel=$\belief$,ylabel=$\objective+\dual\const$,
         x label style={at={(axis description cs:1,-0.05)}},
    y label style={at={(axis description cs:-0.2,1)}},
    width=8cm,
        height=8cm
]
\addplot[color=blue,very thick]coordinates {(0,0) (1/3,0)};
\addplot[color=blue,mark=o] coordinates {(1/3,0)};
\addplot[color=blue,mark=*] coordinates {(1/3,1.05)};
\addplot[color=blue,very thick] coordinates {(1/3,1.05) (2/3,1.05)};
\addplot[color=blue,mark=*] coordinates {(2/3,1.8)};
\addplot[color=blue,mark=o] coordinates {(2/3,1)};
\addplot[color=blue,very thick] coordinates {(2/3,1) (1,1)};
\addplot[color=red,dashed,thick] coordinates {(0,0) (1/3,1.05)};
\addplot[color=red,dashed,thick] coordinates {(2/3,1.8) (1/3,1.05)};
\addplot[color=red,dashed,thick] coordinates {(2/3,1.8) (1,1)};
\end{axis}
\end{tikzpicture}\label{fig:lobbying-high-dual}}
\caption{Lagrangian approach in \autoref{example:lobbying} for $m=0.25$; $\dual\conjugate=\nicefrac{1}{2}$ denotes the optimal multiplier. The dashed red line is the concavification.
}\label{fig:lobbying-lagrangian}
\end{figure}
\end{example}
 
In an inspiring contribution, \cite{le2019persuasion} are the first to highlight the issue raised in \autoref{example:lobbying} in a model with one inequality constraint.
%
%
 At the heart of their result is the observation that the Lagrange multiplier is also part of the solution to the optimization problem. Indeed, they show that the solution corresponds to concavifying a function of $N+1$ variables: the first $N$ correspond to a belief and the last corresponds to the inequality constraint. It follows then that the optimal policy may involve $N+1$ posteriors. The authors also show that the Lagrangian approach is valid for their problem.

Many information design problems involve multiple inequality and equality constraints. For instance, in \cite{doval2020mechanism}, the designer designs both an allocation rule and an information structure; both have to satisfy the agent's participation and incentive compatibility constraints. As another example, consider the problem of persuading a privately informed receiver in \cite{kolotilin2017persuasion}: the designer designs a \emph{menu} of information structures, which has to satisfy the agent's incentive compatibility constraints. Finally, consider the problem of a designer who designs a menu of offers for a privately informed agent, but is limited in how much information the allocation can reveal because of privacy concerns, as in \cite{eilat2021bayesian}. 

We extend the results in \cite{le2019persuasion} to the case of multiple equality and inequality constraints. \autoref{theorem:tomalalemma} shows that the information design problem subject to constraints is equivalent to the solution of a \emph{standard}, but higher dimensional, Bayesian persuasion problem, where the dimensions represent the number of states together with the number of constraints. We use this to derive an upper bound on the number of posteriors induced in an optimal posterior distribution (\autoref{cor:bound}): An optimal posterior distribution induces at most $N+\inequality+\equality$ posteriors, where $N$ is the number of states, \inequality\ is the number of inequality constraints, and $\equality$ is the number of equality constraints. \autoref{cor:tomalalemmadrop} then shows that this upper bound can be refined whenever a constraint does not bind. We also show that the Lagrangian approach is valid. Indeed, \autoref{prop:lagrangian} shows that the constrained information design program can be cast as a Lagrangian where the optimal distribution over posteriors concavifies the Lagrangian at the prior, while the Lagrange multiplier is chosen to minimize the value of this concavification. In other words, the value of the constrained information design problem corresponds to the concavification of the Lagrangian for \emph{some} multiplier. \autoref{theorem:multiplier-existence} shows that a Lagrange multiplier exists under a standard \emph{Slater's condition} (\hyperlink{assumptioncs}{Assumption S}). Examples \ref{example:news} and \ref{example:oo} illustrate how \autoref{prop:lagrangian} can be used to solve constrained information design problems, even without necessarily solving for the optimal multiplier.

Propositions \ref{prop:agreement} and \ref{prop:gid} refine the upper bound on the number of posteriors at an optimal solution that follows from \autoref{cor:bound}. \autoref{prop:agreement} provides an \emph{agreement} condition between the objective and the constraints under which an optimal posterior distribution induces no more posteriors than the number of states. Intuitively, the condition precludes settings like the one in \autoref{example:lobbying}, where the objective function favors information disclosure, while the constraints favor no disclosure. Instead, \autoref{prop:gid} distills and generalizes in the space of posterior beliefs the property that leads to the \emph{recommendation} principle in information design, that is, the result that allows one to equate induced posteriors to action recommendations \citep{myerson1982optimal,kamenica2011bayesian}. Under the conditions of \autoref{prop:gid}, optimal posterior distributions may induce less posteriors than the number of states plus constraints, as we illustrate in \autoref{example:oo}, while at the same time inducing more posteriors than the number of states, as we illustrate in \autoref{example:sm}.


\autoref{sec:applications} shows how \autoref{theorem:tomalalemma} can be leveraged to obtain useful results in two important settings that a priori do not seem to fit the structure of the constrained information design problem introduced in \autoref{sec:main}:

\autoref{sec:limited-commitment} considers the problem of mechanism design with limited commitment. In this application, the set of states of the world corresponds to the agent's private information. \cite{doval2020mechanism} show that it is without loss of generality to consider mechanisms in which the designer designs both an information structure and an allocation. Furthermore, the mechanism must satisfy the agent's participation and incentive compatibility constraints. \autoref{theorem:tomalalemma} implies that it is without loss of generality to focus on mechanisms that induce information structures with finite support and provides an upper bound on the number of posteriors induced by the mechanism. \autoref{prop:med} shows that when the agent's payoff satisfies a version of single-crossing for lotteries (\citealp{bester2007contracting,celik2015implementation,kartik2017single}) the upper bound implied by \autoref{theorem:tomalalemma} can be further reduced (\autoref{cor:bound-limited-commitment-1}).  The assumption of transferable utility provides another way in which this bound can be reduced: When the optimal mechanism can be obtained by maximizing the virtual surplus, the information structure associated to the optimal mechanism uses at most as many posteriors as the number of states of the world (\autoref{prop:vs}). 

\autoref{sec:informed-receiver} considers the problem of persuading a privately informed receiver (\citealp{kolotilin2017persuasion,guo2019interval,candogan2021optimal}). Here the information designer designs a menu of information structures subject to the incentive compatibility constraints of the agent. We show how the designer's problem can be separated into different problems, one for each type of the receiver.\footnote{\cite{candogan2021optimal} make a similar observation in their problem.} We use this decomposition and \autoref{theorem:tomalalemma} to derive an upper bound on the number of posteriors employed in an optimal experiment. Since we make no assumption on the cardinality of the set of receiver actions, the bounds in \autoref{prop:informed-receiver} are the most useful when the set of actions is larger than the set of types.

\paragraph{Related Literature:} The paper builds and expands on the results in \cite{le2019persuasion}. Given the prevalence of \emph{constrained} information design this simple extension is bound to be useful to other researchers. Furthermore, we provide novel applications where these results greatly simplify the analysis. 

Since the first circulation of our draft (see, \citealp{doval2018constrained}), there has been renewed interest in providing tools to solve constrained (information) design problems. \cite{dworczak2019persuasion} apply our results in their study of duality in Bayesian persuasion. \cite{kang2020markets} provides a set of tools complementary to the ones in this paper, by combining results in \cite{bauer1958maximum} and \cite{szapiel1975points}. The results in \cite{szapiel1975points} speak to the set of extreme points in the space of posterior distributions, an infinite-dimensional space, whereas the space of posterior beliefs is finite-dimensional when the set of states is finite. \cite{babichenko2020bayesian} obtain the upper bound in \autoref{cor:bound} using infinite dimensional linear programming tools. They complement the results in our paper by providing computational complexity results. \cite{azrieli2021constrained} illustrates the difference between unconstrained rational inattention problems and those subject to a capacity constraint.

In the context of mechanism design with limited commitment, \cite{bester2007contracting} provide analogues of Propositions \ref{prop:vs} and \ref{prop:postmon}, using tools of infinite dimensional linear programming. While this allows them to conclude that mechanisms in their paper use finitely many output messages, \cite{bester2007contracting} do not provide a characterization of the set of output messages. Therefore, in order to characterize an optimal mechanism, the analyst still has to identify the optimal message space. Instead, we leverage the characterization in \cite{doval2020mechanism}, which allows us to equate the message space of the mechanism to the set of beliefs the designer holds about the agent's type. We then use our results, which are based on the tools of convex analysis employed in the information design literature, to derive an upper bound on the number of posteriors the principal uses at an optimal mechanism.\footnote{\cite{salamanca2021value} studies communication equilibria in sender-receiver games, using a Lagrangian approach to study the sender optimal communication equilibrium. This allows him to derive an upper bound like the one in \autoref{cor:bound} in the context of his model.}

\section{The constrained information design problem}\label{sec:model}
\paragraph{Primitives}  Let \States\ denote a finite set of states, $\States=\{\state_1,\dots,\state_N\}$, and let \Posteriors\ denote the set of probability measures on \States, a subset of the Euclidean space $\reals^\States$.\footnote{To make the comparison with \cite{le2019persuasion} simple, we follow their notation as much as possible. However, while they present their results for any convex set $X$, to make the presentation closer to information design, we let $X$ be the space of beliefs over the set of states \States.}\fnsep\footnote{We impose some technical restrictions on our model. We assume \States\ is Polish, that is a completely metrizable, separable, topological space. We endow it and all Polish spaces in the paper with their Borel $\sigma$-algebra. For a Polish space $X$, we let $\Delta(X)$ denote the set of all Borel probability measures on $X$, endowed with the weak\conjugate\ topology. Thus, $\Delta(X)$ is also a Polish space (Theorem 15.11 in \citealp{aliprantisborder}).\label{ftn:weak-star}} For each element $\prior\in\Posteriors$ and each set $\set\subseteq\Posteriors$, we denote the set of probability distributions on \set\ with mean \prior\ by $\Delta_{\prior}(\set)$. Below we denote by \bsplit\ a distribution over posteriors, and by $\supp\;\bsplit$ and $|\supp\;\bsplit|$ the support and the cardinality of the support of \bsplit, respectively.

We are given a tuple of measurable functions $\left(\objective,\const_1,\dots,\const_\inequality,\dots,\const_{\inequality+\equality}\right):\Posteriors\mapsto\reals^{N+\inequality+\equality}$. Below, the function \objective\ plays the role of the objective function, the tuple $\vectorconstraint_\inequality\equiv(\const_1,\dots,\const_\inequality)$ denotes the functions that correspond to the $\inequality$ inequality constraints, and the tuple $\vectorconstraint_\equality\equiv\left(\const_{\inequality+1},\dots,\const_{\inequality+\equality}\right)$ denotes the functions that correspond to the \equality\ equality constraints, where $\inequality,\equality\geq0$. If $\inequality=0$, then there are no inequality constraints; similarly, if $\equality=0$, then there are no equality constraints. 

\paragraph{The program:} The results of the paper provide tools to analyze the solution to the following program, where $(\prior,\side)\in\mathbb{R}^{N+\inequality+\equality}$:\label{page-opt}
\begin{align}\tag{\OPT}\label{eq:opt}
\Value(\prior,\side)&=\sup_{\bsplit\in\Delta_{\prior}\left(\Posteriors\right)}\mathbbm{E}_\bsplit\left[\objective(\belief)\right]\\
\text{s.t.}&\left\{\begin{array}{l}
\mathbbm{E}_\bsplit\left[\vectorconstraint_\inequality(\belief)\right]\geq\side_\inequality\\
\mathbbm{E}_\bsplit\left[\vectorconstraint_\equality(\belief)\right]=\side_\equality
\end{array}\right.,\nonumber
\end{align}
where the notation $\mathbbm{E}_\bsplit\left[\vectorconstraint_\inequality(\belief)\right]\geq\side_\inequality$ signifies that the vector $\mathbbm{E}_\bsplit\left[\vectorconstraint_\inequality(\belief)\right]\in\reals^\inequality$ is component-wise greater or equal to the vector $\side_\inequality$. Similarly, the notation $\mathbbm{E}_\bsplit\left[\vectorconstraint_\equality(\belief)\right]=\side_\equality$ signifies that the vector $\mathbbm{E}_\bsplit\left[\vectorconstraint_\equality(\belief)\right]\in\reals^\equality$ is component-wise equal to the vector $\side_\equality$. 


\ref{eq:opt} generalizes the constrained information design problem in \cite{le2019persuasion} in two dimensions. First, we allow for both equality and inequality constraints, whereas they consider the case of one inequality constraint ($\inequality=1$) and no equality constraints ($\equality=0$). Second, we define the choice set to be any (Borel) distribution over posteriors with mean \prior, and not just distributions with finite support. As we show in \autoref{prop:finite-support}, restricting attention to distributions with finite support is without loss of generality.

When there are no inequality or equality constraints ($\inequality=\equality=0$), \ref{eq:opt} coincides with the problem in  \cite{kamenica2011bayesian}. \cite{kamenica2011bayesian} show that the value of \ref{eq:opt} coincides with the \emph{concave hull} of the function \objective\ at the prior, \prior. Because the concave hull operator plays an important role in what follows, we define it next:

\begin{definition}[Concave hull of \genericf]\label{definition:concave-hull} Given a function $\genericf:\reals^\dimension\mapsto\reals$, the concave hull of \genericf, denoted \cav\;\genericf, is the function $\cav\space\space\genericf:\reals^\dimension\mapsto\reals$ defined as
\begin{align}\label{eq:cav-hull-h}\tag{\cav\;\genericf}
\left(\cav\;\genericf\right)(\variable)=\sup\left\{\level\in\reals|(\variable,\level)\in\conv\left(\hypograph\;\genericf\right)\right\},
\end{align}
where $\conv\left(\hypograph\;\genericf\right)$ denotes the convex hull of the hypograph of \genericf,  that is, the convex hull of the set $\{(\variable,\level)\in\reals^{\dimension+1}:\level\leq\genericf(\variable)\}$.
\end{definition}
That is, \cav\,\genericf\ is the smallest concave function that majorizes \genericf\ (page 36 in \citealp{rockafellar2015convex}).



Say that \ref{eq:opt} is \emph{feasible} at (\prior,\side) if $\bsplit\in\Delta_{\prior}\left(\Posteriors\right)$ exist such that the constraints in \ref{eq:opt} are satisfied. In what follows, it is useful to distinguish the set of parameters (\prior,\side) for which \ref{eq:opt} is feasible:
\begin{definition}[Feasibility]\label{definition:feasibility}
The set of parameters for which \ref{eq:opt} is feasible is given by
\begin{align}
\feasible=\left\{(\belief,\sideindex)\in\Posteriors\times\reals^{\inequality+\equality}\left|(\exists\bsplit\in\Delta_{\belief}\left(\Posteriors\right)\right.:\begin{array}{l}
\mathbbm{E}_\bsplit\left[\vectorconstraint_\inequality(\belief)\right]\geq\sideindex_\inequality,\;
\mathbbm{E}_\bsplit\left[\vectorconstraint_\equality(\belief)\right]=\sideindex_\equality
\end{array}\right\}.\nonumber
\end{align}
\end{definition}

\paragraph{Assumptions:} We collect here the assumptions that we maintain throughout the analysis and the assumptions we invoke from time to time. To state the first assumption, recall that the \emph{effective domain} of a function \genericf\ on $\reals^\dimension$ is the set $\dom\;\genericf=\{\variable\in\reals^D:\genericf(\variable)>-\infty\}$.
\begin{assumption}[Maintained assumptions on \objective\ and \vectorconstraint\ ]\label{assumption:main} The effective domain of (\objective,\vectorconstraint) is the set \Posteriors. Furthermore, $\objective,\vectorconstraint<+\infty$ for all $\belief\in\Posteriors$.
%
\end{assumption}
The next assumption is a standard requirement on the objective function in the information design literature:
\begin{assumptionf}\hypertarget{assumptionf}{} The objective function \objective\ is upper-semicontinuous.
\end{assumptionf}
The last assumption plays a similar role to \hyperlink{assumptionf}{Assumption \objective} for the constraint functions \vectorconstraint. We invoke it below to infer properties of the set \feasible:
\begin{assumptiong}\hypertarget{assumptiong}{} $\vectorconstraint_\inequality$ is component-wise upper-semicontinuous, while $\vectorconstraint_\equality$ is component-wise continuous.
\end{assumptiong}
We note the following two implications of our assumptions. First, \hyperlink{assumptiong}{Assumption $\vectorconstraint$} implies that the set \feasible\ is closed. 
Second, Assumptions \hyperlink{assumptionf}{\objective} and \hyperlink{assumptiong}{\vectorconstraint} imply that if $(\prior,\side)\in\feasible$, then the supremum in \ref{eq:opt} is attained. We record this below and prove it in the supplementary material \citep{supplement}:
\begin{observation}[\feasible\ is closed]\label{observation:f-closed} Under \hyperlink{assumptiong}{Assumption $\vectorconstraint$}, \feasible\ is closed.
\end{observation}
\begin{observation}[\ref{eq:opt} has a solution]\label{observation:solution} Suppose $(\prior,\side)\in\feasible$ and Assumptions \hyperlink{assumptionf}{\objective} and \hyperlink{assumptiong}{\vectorconstraint} hold. Then, the value of \ref{eq:opt}, \Value(\prior,\side), is attained.
\end{observation}

\section{Main results}\label{sec:main}
Our first main result, \autoref{theorem:tomalalemma}, connects the value of \ref{eq:opt} to the concave hull of a \emph{modified} version of the objective function \objective, which we introduce next. 

\paragraph{A modified objective function:}  Let $\constraintset=\{(\belief,\sideindex)\in\Posteriors\times\reals^{\inequality+\equality}:\vectorconstraint_\inequality(\belief)\geq\sideindex_\inequality,\vectorconstraint_\equality(\belief)=\sideindex_\equality\}$. That is, \constraintset\ is the subset of $\Posteriors\times\reals^{\inequality+\equality}$ which satisfies a \emph{pointwise} version of the constraints in \ref{eq:opt}. To see this, contrast the set \constraintset\ with \feasible. Whereas \ref{eq:opt} is feasible at $(\belief,\sideindex)$ if a distribution with mean \belief\ exists that satisfies the constraints at \sideindex\ \emph{in expectation}, $(\belief,\sideindex)\in\constraintset$  if the distribution that places probability $1$ on \belief\ satisfies the constraints at \sideindex.
%

Given \constraintset, define the function $\modifiedobjective:\reals^{N+\inequality+\equality}\mapsto\reals\cup\{\pm\infty\}$ as follows
\begin{align}\label{eq:modified-objective}
\modifiedobjective(\belief,\sideindex)=\objective(\belief)-\delta(\belief,\sideindex|\constraintset),
\end{align}
where $\delta(\belief,\sideindex|\constraintset)$ is the indicator function of \constraintset, taking value $0$ if $(\belief,\sideindex)\in \constraintset$ and $+\infty$ otherwise \citep{rockafellar2015convex}.

\autoref{theorem:tomalalemma}, relates the solution to \ref{eq:opt} to the concave hull of the $N+\equality+\inequality$-dimensional function $\modifiedobjective$:
\begin{theorem}\label{theorem:tomalalemma}
For each $(\prior,\side)\in\reals^{N+\inequality+\equality}$, the value of \ref{eq:opt} at (\prior,\side) coincides with the value of the concave hull of \modifiedobjective\ at (\prior,\side). That is, 
\begin{align}
\Value(\prior,\side)=\cav\;\modifiedobjective(\prior,\side).
\end{align}
\end{theorem}
\autoref{theorem:tomalalemma} characterizes in closed form the value function for \ref{eq:opt} via the concave hull of the modified objective function, \modifiedobjective. Key to this result is \autoref{lemma:convex-domain}, where we show that the (effective) domain of \cav\;\modifiedobjective\ is the set of parameters for which \ref{eq:opt} is feasible, \feasible. In other words, the convex hull of the set \constraintset\ is the set \feasible. Thus, while \modifiedobjective\ is defined relative to the more stringent requirement that the constraints are satisfied \emph{pointwise}, the domain of its concave hull is the set \feasible. Within this domain, that is, within the constraints defined by \ref{eq:opt}, standard ``information design logic'' implies that \cav\;\modifiedobjective(\prior,\side) describes the largest value that $\mathbb{E}_\bsplit[\objective]$ can attain.


\begin{example}[\autoref{example:lobbying} continued]
\autoref{fig:lobbying-theorem} illustrates \autoref{theorem:tomalalemma} in the context of \autoref{example:lobbying}. \autoref{fig:lobbying-domain} shows the set \constraintset\ in red; the areas in blue and red depict its convex hull, \feasible. In the case of one inequality constraint, the set \feasible\ coincides with the convex hull of the hypograph of the function \vectorconstraint, so that the upper boundary of the set depicted in \autoref{fig:lobbying-domain} is $\cav\;\vectorconstraint$, for \vectorconstraint\ as in \autoref{eq:lobbying-constraint}.  This is intuitive: \ref{eq:opt} is feasible at (\prior,\side) if and only if $\cav\;\vectorconstraint(\prior)$ satisfies the constraint at \side, since this is the largest value that $\mathbb{E}_\bsplit\vectorconstraint$ may attain. \autoref{fig:lobbying-domain} also shows that whenever the value of the constraint, \side, is strictly positive, then \ref{eq:lobbying-opt} is not feasible for low enough and large enough values of the prior. The reason is that for priors $\prior\in[0,\nicefrac{\side}{3})\cup(1-\nicefrac{\side}{3},1]$,  no Bayes' plausible posterior distribution exists that assigns enough probability to the event $\belief\in[\nicefrac{1}{3},\nicefrac{2}{3}]$.

\autoref{fig:lobbying-3d} depicts the function \modifiedobjective\ for \autoref{example:lobbying}. Computing the concave hull of \modifiedobjective\ and then taking its section at $\side=\sideindex$ for different values of \sideindex, we obtain \autoref{fig:lobbying-cav}. \autoref{fig:lobbying-cav} depicts in dashed red the function $\cav\;\modifiedobjective(\cdot,0)$: When \side=0, the constraint is always satisfied, so that $\cav\;\modifiedobjective(\cdot,0)$ coincides with \cav\;\objective. Similarly, in black we have $\cav\;\modifiedobjective(\cdot,0.75)$. As explained above, when $\prior<1/4$ or $\prior>3/4$, \ref{eq:lobbying-opt} is not feasible, so that $\cav\;\modifiedobjective(\cdot,0.75)\equiv-\infty$. Instead, when $\prior\in[1/2,3/4]$, the constraint does not bind, so that $\cav\;\modifiedobjective(\cdot,0.75)=\cav\;\objective$, whereas when $\prior\in[0.25,0.5)$, the constraint does bind and $\cav\;\modifiedobjective(\cdot,0.75)$ is strictly below the unconstrained concave hull of \objective\ (dashed red).

\begin{figure}[th!]
\subfloat[The set \constraintset\ (red) and its convex hull \feasible\ (red$+$blue)]{
\begin{tikzpicture}[scale=0.8]
\begin{axis}[axis lines=middle, xmin=-0.05,xmax=1.1,ymin=-4,ymax=1.2,xlabel=$\belief$,xtick={0,1/3,2/3,1},xticklabels={0,$\frac{1}{3}$,$\frac{2}{3}$,1}, ylabel=$\sideindex$, y label style={at={(axis description cs:-0.1,1)}}, x label style={at={(axis description cs:1,0.7)}},
    width=8cm,
        height=8cm
]
\addplot[draw=red,thick]coordinates{(0,0) (0,-4)};
\addplot[draw=red,thick]coordinates{(1,0) (1,-4)};
\addplot[draw=red,thick]coordinates{(0,0) (1/3,0)};
\addplot[draw=red,thick]coordinates{(1/3,0) (1/3,1)};
\addplot[draw=red,thick]coordinates{(1/3,1) (2/3,1)};
\addplot[draw=red,thick]coordinates{(2/3,0) (2/3,1)};
\addplot[draw=red,thick]coordinates{(2/3,0) (1,0)};
\addplot[thick,draw=none,fill=red,
                    fill opacity=0.05]coordinates {
            (0, 0) 
            (0,-4)
            (1/3,-4) (1/3, 0)  };
\addplot[thick,draw=none,fill=red,
                    fill opacity=0.05]coordinates {
            (1/3, 1) 
            (1/3,-4)
            (2/3,-4) (2/3, 1)  };
\addplot[thick,draw=none,fill=red,
                    fill opacity=0.05]coordinates {
            (2/3, 0) 
            (2/3,-4)
            (1,-4) (1, 0)  };
            
            \addplot[draw=blue,thick]coordinates{(0,0) (1/3,1)};
                        \addplot[draw=blue,thick]coordinates{(1,0) (2/3,1)};
\addplot[thick,draw=none,fill=blue,
                    fill opacity=0.05]coordinates {
            (0, 0) 
            (1/3,1)
            (1/3,0) (0, 0)  };
\addplot[thick,draw=none,fill=blue,
                    fill opacity=0.05]coordinates {
            (2/3, 1) 
            (1,0)
            (2/3,0) (2/3, 1)  };
\end{axis}
\end{tikzpicture}\label{fig:lobbying-domain}}
\subfloat[\modifiedobjective\ on its effective domain]{\begin{tikzpicture}[scale=0.8]
\begin{axis}[axis lines=middle, xmin=-0.05,xmax=1.1,ymin=-4,ymax=1.2,,zmin=-0.05,zmax=1,xlabel=$\belief$, ylabel=$\sideindex$,  zlabel=$\modifiedobjective$,
    width=8cm,
        height=8cm
]
\addplot3[draw=none,fill=red,fill opacity=0.25]coordinates{(0,0,0) (1/3,0,0) (1/3,-4,0) (0,-4,0)};
\addplot3[draw=none,fill=red,fill opacity=0.25]coordinates{(1/3,1,0.25) (2/3,1,0.25) (2/3,-4,0.25) (1/3,-4,0.25)};
\addplot3[draw=red,thick]coordinates{(1/3,1,0.25) (1/3,-4,0.25)};
\addplot3[draw=red,thick]coordinates{(2/3,0,1) (2/3,-4,1)};
\addplot3[draw=none,fill=red,fill opacity=0.25]coordinates{(2/3,0,1) (1,0,1) (1,-4,1) (2/3,-4,1)};
\end{axis}
\end{tikzpicture}\label{fig:lobbying-3d}}
\subfloat[$\cav\;\modifiedobjective(\cdot,\side)$ for different values of \side.]{\begin{tikzpicture}[scale=0.8]
\begin{axis}[axis lines=middle, xmin=-0.05,xmax=1.1,ymin=-0.05,ymax=1,xlabel=$\belief$, ylabel=$\cav\;\modifiedobjective$, xtick={0.25,0.5,2/3,0.75},xticklabels={$\frac{1}{4}$,$\frac{1}{2}$,$\frac{2}{3}$,$\frac{3}{4}$},
ytick={0.5,1}, yticklabels={0.5,1}, y label style={at={(axis description cs:-0.15,1)}}, x label style={at={(axis description cs:1,-0.05)}},
    width=8cm,
        height=8cm,legend style={fill=none, at={(0.8,0.2)},anchor=west},legend cell align={left}]
]

\addplot[draw=red,dashed,thick]coordinates{(0,0) (2/3,1)};
\addlegendentry{$\side=0$};
\addplot[draw=black,thick,domain=0.25:0.5]{(9/4)*x-(3/8)};
\addlegendentry{$\side=0.75$};
\addplot[draw=blue,dashed,thick,domain=0.3:0.6]{(9/4)*x-(9/20)};
\addlegendentry{$\side=0.9$};
\addplot[draw=red,dashed,thick]coordinates{(1,1) (2/3,1)};

\addplot[draw=black,thick] coordinates {(1/2,0.75) (2/3,1)};
\addplot[draw=black,thick] coordinates {(2/3,1) (3/4,1)};


\addplot[draw=blue,dashed,thick] coordinates {(0.6,0.9) (2/3,1)};
\addplot[draw=blue,dashed,thick] coordinates {(2/3,1) (3/4,1)};
\addplot[draw=black,loosely dashed] coordinates {(0.25,0) (0.25,3/16)};
\addplot[draw=black,loosely dashed] coordinates {(0.5,0) (0.5,3/4)};
\addplot[draw=black,loosely dashed] coordinates {(2/3,0) (2/3,1)};
\end{axis}
\end{tikzpicture}\label{fig:lobbying-cav}
}

\caption{The modified objective function \modifiedobjective\ in \autoref{example:lobbying}.}\label{fig:lobbying-theorem}
\end{figure}
\end{example}

\paragraph{Bounding the number of posteriors at solutions of \ref{eq:opt}:} By relating the value function of \ref{eq:opt} to the concavification of the $N+\inequality+\equality$-dimensional function, \modifiedobjective, \autoref{theorem:tomalalemma} together with Corollary 17.1.5 in \cite{rockafellar2015convex} imply the following property of solutions to \ref{eq:opt}, whenever the supremum in \ref{eq:opt} is attained:\label{page-ct-cor-cor}\footnote{Theorem 4.8 in \cite{anderson1987linear} provides an analogue to \autoref{cor:bound} under the assumption that the objective and constraint functions are continuous. Instead, we rely on the result in \cite{rubin1958note} to show that restricting the choice set in \ref{eq:opt} to finite-support distributions is without loss and then on \autoref{theorem:tomalalemma} to derive the upper bound using the definition of the \cav\ operator (see \autoref{prop:conv-hull-new} in \autoref{appendix:main}).}
\begin{cor}\label{cor:bound}
Suppose $(\prior,\side)\in\feasible$ and the supremum in \ref{eq:opt} is attained. Then,  a solution $\bsplitopt$ to \ref{eq:opt} exists with $|\supp\;\bsplitopt|\leq N+\inequality+\equality$. 
\end{cor}
\autoref{example:lobbying} shows that the second bound is tight. Furthermore, we can relate the upper bound on the number of posteriors at an optimal solution to the number of binding constraints:
\begin{cor}\label{cor:tomalalemmadrop}
Suppose $(\prior,\side)\in\feasible$, the supremum in \ref{eq:opt} is attained, and that only $\binding<\inequality$ inequality constraints bind. Then, a solution $\bsplitopt$ to \ref{eq:opt} exists with $|\supp\;\bsplitopt|\leq N+\binding+\equality$. 
\end{cor}
We close this part with two remarks about \autoref{theorem:tomalalemma}. First, \autoref{theorem:tomalalemma} provides an alternative way to see why Assumptions \hyperlink{assumptionf}{\objective} and \hyperlink{assumptiong}{\vectorconstraint} imply that a solution to \ref{eq:opt} exists whenever $(\prior,\side)\in\feasible$. Under those assumptions, it follows that \modifiedobjective\ is upper-semicontinuous, by virtue of being the sum of two upper-semicontinuous functions, \objective, and the negative of the indicator of the closed set, \constraintset. It then follows that under those assumptions the value of \ref{eq:opt} is attained in the set \feasible. Second, whereas \autoref{theorem:tomalalemma} is written in terms of the optimization problem \ref{eq:opt}, the proof actually shows that the convex hull of the graph of \modifiedobjective\ coincides with the convex hull of the graph of the function $\jointf=(\objective,\vectorconstraint)$ over the set \constraintset. We use this observation when we show that a Lagrange multiplier exists for \ref{eq:opt} in \autoref{theorem:multiplier-existence}.

\subsection{Validating the Lagrangian approach}
By delivering a closed form solution to the value of \ref{eq:opt} in terms of the \cav\ operator, \autoref{theorem:tomalalemma} provides a transparent way to understand why concavification subject to constraints may necessitate solutions \bsplit\ whose support contains more posteriors than the number of states. However, outside of simple examples like \autoref{example:lobbying}, solving for \cav\;\modifiedobjective\ is not necessarily tractable, even with a binary state space. There are at least three related reasons. First, the domain of \modifiedobjective\ always has dimension larger than $N$ and it increases with each additional constraint. Second, solving for \cav\;\modifiedobjective\ implies solving for a distribution over $(\posterior,\sideindex)$ with mean (\prior,\side), when we are ultimately interested in the marginal of this distribution over \Posteriors. Third, as illustrated in \autoref{fig:lobbying-cav}, it is important to solve for the \emph{joint} distribution over (\posterior,\sideindex) to capture the points at which \ref{eq:opt} is feasible. To see this, note first that the effective domain of $\cav\;\modifiedobjective(\cdot,\side)$ depends on \side. Moreover, as evidenced by the difference between \cav\;\objective\ and $\cav\;\modifiedobjective(\cdot,\side)$, the beliefs \posterior\ used to attain the latter may also depend on the value of \side.

As we illustrate throughout the paper and is evidenced by the body of work in constrained information design, the Lagrangian approach maybe more efficient. After all, no matter the number of constraints, the Lagrangian objective is an $N$-dimensional function whose effective domain is the set of posterior beliefs \Posteriors. Thus, when concavifying the Lagrangian, one can rely on the standard tools of information design (see, for instance, \autoref{observation:visual} below). Thus, understanding under what conditions a Lagrangian approach is valid is definitely of use.  Our second main result, \autoref{prop:lagrangian}, validates the Lagrangian approach:
\begin{theorem}\label{prop:lagrangian}
Suppose that \ref{eq:opt} is feasible at (\prior,\side). Then, the following holds
\begin{align}\tag{WD}\label{eq:weak-duality-cav}
\Value(\prior,\side)\leq\inf_{t\in\reals_+^\inequality\times\reals^\equality}\left[\cav\left(\objective+\sum_{k=1}^{\inequality+\equality}\dual_k\const_k\right)(\prior)-\sum_{k=1}^{\inequality+\equality}\dual_k\side_k\right].
\end{align}
Furthermore, if $(\prior,\side)\in\interior\;\feasible$, then the following holds
\begin{align}\tag{NDG}\label{eq:no-duality-gap}
\Value(\prior,\side)=\inf_{t\in\reals_+^\inequality\times\reals^\equality}\left[\cav\left(\objective+\sum_{k=1}^{\inequality+\equality}\dual_k\const_k\right)(\prior)-\sum_{k=1}^{\inequality+\equality}\dual_k\side_k\right].
\end{align}
Finally, if $\feasible\neq\emptyset$ and Assumptions \hyperlink{assumptionf}{\objective} and \hyperlink{assumptiong}{\vectorconstraint} hold, then \ref{eq:no-duality-gap} holds for all $(\prior,\side)\in\feasible$.
\end{theorem}
\autoref{prop:lagrangian} clarifies the sense in which the Lagrangian approach is valid for \ref{eq:opt}. \autoref{eq:no-duality-gap} states that we can interpret the value of \ref{eq:opt} as obtaining from the following procedure: For each candidate Lagrange multiplier \dual, we obtain the concave closure of the Lagrangian \emph{at the prior \prior} and then we choose the Lagrange multiplier \dual\ to minimize the Lagrangian. 

\autoref{theorem:tomalalemma} relates the value of \ref{eq:opt} to the concave hull of the modified objective function \modifiedobjective, whereas the proof of \autoref{prop:lagrangian} relates the concave closure of \modifiedobjective\ to the corresponding Lagrangian of \ref{eq:opt}. Thus, in general, we only obtain an upper bound on the value of \ref{eq:opt} as in \autoref{eq:weak-duality-cav}.
 Now, the proof of \autoref{theorem:tomalalemma} implies that the effective domain of \cav\;\modifiedobjective\ is \feasible. Because the concave closure and the concave hull of \modifiedobjective\ coincide in the interior of their effective domain (Theorem 7.4 in \citealp{rockafellar2015convex}), it then follows that in the interior of \feasible, the value of \ref{eq:opt} coincides with that of the optimized Lagrangian, which leads to \ref{eq:no-duality-gap}.\footnote{\autoref{prop:lagrangian} provides the analogue to Theorem 3.3, item 2 in \cite{le2019persuasion}, while clarifying the assumptions for their result to hold. In particular, in their proof, the authors equate the concave hull of \modifiedobjective\ with its concave closure. Without further assumptions on \modifiedobjective\ or the constraint set \constraintset, Theorem 7.4 in \cite{rockafellar2015convex} implies that these functions coincide everywhere except on the relative boundary of \feasible\ (see \autoref{example:cav-not-clcav}). This motivates the restriction to the interior of \feasible\ in the statement of \autoref{prop:lagrangian}. Furthermore, under Assumptions \hyperlink{assumptionf}{\objective} and \hyperlink{assumptiong}{\vectorconstraint}, \cav\;\modifiedobjective\ is  upper-semicontinuous and hence, coincides with the concave closure of \modifiedobjective\ everywhere.}

As we show in the supplementary material \citep{supplement}, \autoref{prop:lagrangian} states in the language of concavification that there is no duality gap between \ref{eq:opt} and its dual in the interior of \feasible.\footnote{\ref{eq:opt} is an instance of the dual of a semi-infinite linear program \citep{anderson1987linear}. Assuming that the objective and constraint functions are continuous, Theorem 4.4 in \cite{anderson1987linear} provides conditions under which the value of the dual coincides with the value of the primal. For instance, when $(\prior,\side)\in\interior\;\feasible$, then the conditions of item (b) of Theorem 4.4 hold, since by \autoref{assumption:main}, $\Value(\prior,\side)<+\infty$ for $(\prior,\side)\in\feasible$.\label{ftn:duality}}  Indeed, we show that the right hand side of \ref{eq:no-duality-gap} is the value of the dual to \ref{eq:opt}. \label{page-duality} In other words, the first part of the statement of \autoref{prop:lagrangian} simply says that weak duality holds for \ref{eq:opt}, whenever it is feasible at (\prior,\side).

However, by restating the dual in terms of the concavification of an $N$-dimensional function, $\objective+\sum_{k=1}^{\inequality+\equality}\dual_k\const_k$, \autoref{prop:lagrangian} becomes useful in applications. Indeed, the following observation is standard in information design:
\begin{observation}\label{observation:visual}
Suppose \bsplit\ attains $\cav(\objective+\sum_{k=1}^{\inequality+\equality}\dual_k\const_k)(\prior)$. Then, 
\begin{align}
\support\;\tau\subseteq\left\{\belief\in\Posteriors:\cav(\objective+\sum_{k=1}^{\inequality+\equality}\dual_k\const_k)(\belief)=(\objective+\sum_{k=1}^{\inequality+\equality}\dual_k\const_k)(\belief)\right\},
\end{align}
that is \bsplit\ only induces posteriors for which there is no value to persuasion.
\end{observation}
We illustrate the usefulness of \autoref{prop:lagrangian} together with \autoref{observation:visual} in \autoref{example:news}:\footnote{For an example in the literature that also relies on \autoref{observation:visual} to arrive at an optimal solution through the concavification of the Lagrangian, see \cite{rosar2017test}.}
\begin{example}[Newsroom]\label{example:news}A news platform designs how information is released to consumers and wishes to be perceived as unbiased. We represent this as follows. There are two equally likely states of the world, $\States=\{\state_L,\state_R\}$. The platform's payoff is given by
\begin{align*}
\objective(\belief)=\left|\frac{1}{2}-\belief\right|+0.4,
\end{align*}
where $\belief\in[0,1]$ is the likelihood a consumer who gets their news on the platform assigns to the true state being $\state=\state_R$ after getting the news from the platform. Consistent with its desire to provide unbiased content, the platform's payoffs are maximized when consumers learn the true state.

In order to operate, the platform must be able to collect ad revenue and for this it requires a broad audience. A left (right) leaning consumer enjoys reading the news on the platform whenever the platform confirms the consumer's views of the world, which we model by the following payoff functions:
\begin{align*}
\const_L(\belief)&=\left\{\begin{array}{ll}0.4&\text{ if }\belief\leq0.6\\
1.3-1.5\belief&\text{ if } 0.6<\belief\leq0.8\\
0.5-0.5\belief&\text{ if }0.8<\belief\leq1\end{array}\right.,&
\const_R(\belief)&=\left\{\begin{array}{ll}0.4&\text{ if }\belief\geq0.4\\
1.5\belief-0.2&\text{ if } 0.2\leq\belief<0.4\\
0.5\belief&\text{ if }0\leq\belief<0.2\end{array}\right..
\end{align*}
A left (right) leaning reader gets their news from the platform if her expected payoff is larger than $\side_L$ ($\side_R$), where $\side_\cdot\in(0.2,0.4)$ and $8-5\side_R-12.5\side_L\leq3$. \autoref{fig:news-obj-cons} depicts the platform's and the readers' payoff functions.

\begin{figure}[t!]
\centering
\subfloat[Objective function]{\begin{tikzpicture}[scale=0.8]
\begin{axis}[axis lines=middle,xmin=-0.05,xmax=1.1,ymin=0.35,ymax=1,xtick={0,0.5,1},xticklabels={0,$\prior$,1},ytick={0,0.5,1},xlabel=$\belief$,ylabel=$\objective$,   x label style={at={(axis description cs:1,-0.1)}},
    y label style={at={(axis description cs:-0.05,1)}},
    width=8cm,
        height=8cm]
\addplot[color=black,thick]coordinates {(0,0.9) (0.5,0.4)};
\addplot[color=black,thick]coordinates {(0.5,0.4) (1,0.9)};
\end{axis}\end{tikzpicture}}
\subfloat[Constraints]{\begin{tikzpicture}[scale=0.8]
\begin{axis}[axis lines=middle,xmin=-0.05,xmax=1.1,ymin=-0.05,ymax=0.75,,xtick={0,0.2,0.4,0.5,0.6,0.8,1},xticklabels={0,$\frac{1}{5}$,$\frac{2}{5}$,$\prior$,$\frac{3}{5}$,$\frac{4}{5}$,1},ytick={0,0.5},xlabel=$\belief$,ylabel=$\const_\cdot$,   x label style={at={(axis description cs:1,-0.05)}},
    y label style={at={(axis description cs:-0.05,1)}},
    width=8cm,
        height=8cm,legend style={fill=none}]
\addplot[color=blue,thick]coordinates {(0,0.4) (0.6,0.4)};
\addlegendentry{$\const_L$};
\addplot[color=red,thick,domain=0:0.2]{0.5*x};
\addlegendentry{$\const_R$};
\addplot[color=red,thick,domain=0.2:0.4]{1.5*x-0.2};
\addplot[color=red,thick,domain=0.4:1]{0.4};
\addplot[color=blue,thick]coordinates {(0.6,0.4) (0.8,0.1)};
\addplot[color=blue,thick]coordinates {(0.8,0.1) (1,0)};
\end{axis}\end{tikzpicture}}
\caption{Objective function (left) and constraints (right) in \autoref{example:news}}\label{fig:news-obj-cons}
\end{figure}

Without readers, the platform gets no revenues so the optimal news provision policy solves:
\begin{align}\tag{\OPT$_{N}$}\label{eq:opt-news}
\max_{\bsplit\in\Delta_{\prior}\Posteriors}&\mathbb{E}_\bsplit\left[\objective\right]\\
\text{ s.t. }&\mathbb{E}_\bsplit\left[\const_i\right]\geq\side_i,i\in\{L,R\}.\nonumber
\end{align}
That is, the news platform has lexicographic preferences over ad revenue and its desire to appear neutral. The platform first needs to guarantee ad revenue and hence, ensure readership. Having guaranteed readership, the platform then prefers to provide news with as little slant as possible.

\autoref{prop:lagrangian} implies that in order to solve the platform's problem we can consider the following Lagrangian objective function:
\begin{align}\label{eq:news-lagrangian}
\left(\objective+\dual\const\right)(\belief)&=\left\{\begin{array}{ll}0.9-\belief+\dual_L0.4+\dual_R0.5\belief&\text{ if }\belief\in[0,0.2)\\
0.9-\belief+\dual_L0.4+\dual_R(1.5\belief-0.2)&\text{ if }\belief\in[0.2,0.4)\\
\left|\belief-0.5\right|+0.4(\dual_L+\dual_R)&\text{ if }\belief\in[0.4,0.6)\\
\belief-0.1+\dual_L(1.3-1.5\belief)+\dual_R0.4&\text{ if }\belief\in[0.6,0.8)
\\
\belief-0.1+\dual_L(0.5-0.5\belief)+\dual_R0.4&\text{ otherwise }\end{array}\right..
\end{align}
We now apply \autoref{observation:visual} to \autoref{fig:news-lagrangian}, which depicts the Lagrangian and its concavification at the prior, to derive conclusions about the posterior distributions that achieve that concavification and whether they can satisfy the constraints. For instance, in the case of \autoref{fig:news-low-dual} it is enough to consider posterior distributions that place probability only on $\{0,1\}$, whereas for \autoref{fig:news-high-dual} it is enough to consider distributions whose support lies in $\{0.4,0.6\}$. This automatically rules out that a solution exists where the multipliers, $\dual_L,\dual_R$, are as in \autoref{fig:news-low-dual}, since this would lead to an infeasible solution. Instead, inspection of \autoref{fig:news-opt-dual} shows that optimal posterior distributions would have support in a subset of $\{0,0.4,0.6,1\}$. Indeed, the optimal solution has exactly this support:\footnote{The optimal distribution has support $(0,0.4,0.6,1)$ with weights $(\nicefrac{2}{3}+\nicefrac{5}{6}\side_R-\nicefrac{5}{3}\side_L,1-\nicefrac{5}{2}\side_R,\nicefrac{5}{3}\side_R+\nicefrac{25}{6}\side_L-\nicefrac{5}{3},1-\nicefrac{5}{2}\side_L)$.} the platform balances its desire to appear neutral with enough political content to attract its audience. Similar to \autoref{example:lobbying}, there is another posterior distribution which attains the concavification of the Lagrangian and has binary support; namely, fully revealing the state. However, this is clearly not feasible.

\begin{figure}[t!]
\centering
\subfloat[$\dual_L=\dual_R=0.5<\dual\conjugate$]{
\begin{tikzpicture}[scale=0.8]
\begin{axis}[axis lines=middle,xmin=-0.05,xmax=1.1,ymin=0.6,ymax=1.25,xtick={0,0.2,0.4,0.5,0.6,0.8,1},xticklabels={0,$\frac{1}{5}$,$\frac{2}{5}$,$\prior$,$\frac{3}{5}$,$\frac{4}{5}$,1},ytick={0,0.5,1},xlabel=$\belief$,ylabel=$\objective+\sum_{i\in\{L,R\}}\dual_i\const_i$,   x label style={at={(axis description cs:1,-0.1)}},
    y label style={at={(axis description cs:-0.4,1)}},
    width=8cm,
        height=8cm]
\addplot[thick,domain=0:0.2]{1.1-0.75*x};
\addplot[thick,domain=0.2:0.4]{1-0.25*x};
\addplot[thick,domain=0.4:0.5]{1.3-x};
\addplot[thick,domain=0.5:0.6]{x+0.3};
\addplot[thick,domain=0.6:0.8]{0.25*x+0.75};
\addplot[thick,domain=0.8:1]{0.75*x+0.35};
\addplot[color=red,dashed,thick] coordinates {(0,1.1) (1,1.1)};
\end{axis}
\end{tikzpicture}\label{fig:news-low-dual}}
\subfloat[$\dual_L=\dual_R=\dual\conjugate$]{\begin{tikzpicture}[scale=0.8]
\begin{axis}[axis lines=middle,xmin=-0.05,xmax=1.1,ymin=1.15,ymax=1.35,,xtick={0,0.2,0.4,0.5,0.6,0.8,1},xticklabels={0,$\frac{1}{5}$,$\frac{2}{5}$,$\prior$,$\frac{3}{5}$,$\frac{4}{5}$,1},ytick={0,0.5,1},xlabel=$\belief$,ylabel=$\objective+\sum_{i\in\{L,R\}}\dual_i\conjugate\const_i$,   x label style={at={(axis description cs:1,-0.1)}},
    y label style={at={(axis description cs:-0.4,1)}},
    width=8cm,
        height=8cm]
\addplot[thick,domain=0:0.2]{1.3-0.5*x};
\addplot[thick,domain=0.2:0.4]{1.1+0.5*x};
\addplot[thick,domain=0.4:0.5]{1.7-x};
\addplot[thick,domain=0.5:0.6]{x+0.7};
\addplot[thick,domain=0.6:0.8]{-0.5*x+1.6};
\addplot[thick,domain=0.8:1]{0.5*x+0.8};
\addplot[color=red,dashed,thick] coordinates {(0,1.3) (1,1.3)};
\end{axis}
\end{tikzpicture}\label{fig:news-opt-dual}}

\subfloat[$\dual_L=\dual_R=1.5>\dual\conjugate$]{\begin{tikzpicture}[scale=0.8]
\begin{axis}[axis lines=middle,xmin=-0.05,xmax=1.1,ymin=1,ymax=2,,xtick={0,0.2,0.4,0.5,0.6,0.8,1},xticklabels={0,$\frac{1}{5}$,$\frac{2}{5}$,$\prior$,$\frac{3}{5}$,$\frac{4}{5}$,1},ytick={0,0.5,1},ytick={0,0.5,1},xlabel=$\belief$,ylabel=$\objective+\sum_{i\in\{L,R\}}\dual_i\const_i$,   x label style={at={(axis description cs:1,-0.1)}},
    y label style={at={(axis description cs:-0.4,1)}},
    width=8cm,
        height=8cm]
\addplot[thick,domain=0:0.2]{1.5-0.25*x};
\addplot[thick,domain=0.2:0.4]{1.2+1.25*x};
\addplot[thick,domain=0.4:0.5]{2.1-x};
\addplot[thick,domain=0.5:0.6]{x+1.1};
\addplot[thick,domain=0.6:0.8]{-1.25*x+2.45};
\addplot[thick,domain=0.8:1]{0.25*x+1.25};
\addplot[color=red,dashed,thick] coordinates {(0.4,1.7) (0.6,1.7)};
\addplot[color=red,dashed,thick] coordinates { (0.6,1.7) (1, 1.5)};
\addplot[color=red,dashed,thick] coordinates {(0.4,1.7) (0,1.5)};
\end{axis}
\end{tikzpicture}\label{fig:news-high-dual}}
\caption{Lagrangian approach in \autoref{example:news}; \dual\conjugate\ denotes the optimal multiplier. The dashed red line is the concavification}\label{fig:news-lagrangian}
\end{figure}
\end{example}

\paragraph{Existence of a Lagrange multiplier:} The analysis so far leaves open the question of whether a Lagrange multiplier exists such that the value of \ref{eq:opt} corresponds to the concavification of the Lagrangian at the prior. \autoref{theorem:multiplier-existence} below confirms this is the case under the following \emph{Slater's condition}. To introduce this condition, let $\jointcav$ denote the section at the prior \prior\ of the convex hull of the graph of the function \jointf=(\objective,\vectorconstraint), that is, $\jointcav=\{\level\in\reals^{\inequality+\equality+1}|(\prior,\level)\in\conv\left(\graph\;\jointf\right)\}$. The set \jointcav\ describes the values of the objective and the constraints that are \emph{jointly} feasible under the Bayes' plausibility constraint. That is, $\level\in\jointcav$ if and only if $\bsplit\in\Delta_{\prior}\Posteriors$ exists such that $\level=\mathbb{E}_\bsplit[\jointf]$ (cf., \citealp{boleslavsky2018bayesian,doval2021information}).
\begin{assumptioncs}\hypertarget{assumptioncs}{} \ref{eq:opt} satisfies \emph{Slater's condition} at $(\prior,\side)\in\feasible$ if a point in the interior of \jointcav, $(\level_f,\level_{\inequality+\equality}^\star)$, exists such that $\level_\inequality^\star\geq\side_\inequality$ and $\level_E^\star=\side_\equality$.
\end{assumptioncs}
\begin{theorem}[Existence of a Lagrange multiplier]\label{theorem:multiplier-existence}
Suppose \ref{eq:opt} satisfies \hyperlink{assumptioncs}{Assumption S} at $(\prior,\side)$. Then, a Lagrange multiplier, $\dual\conjugate\in\reals^{\inequality+\equality}$, exists such that $\dual_\inequality\conjugate\geq0$ and 
\begin{align}\label{eq:cav-lagrangian}
\Value(\prior,\side)=\cav\left(\objective+\sum_{k=1}^{\inequality+\equality}\dual_k\conjugate\const_k\right)(\prior)-\sum_{k=1}^{\inequality+\equality}\dual_k\conjugate\side_k.
\end{align}
\end{theorem} 
The proof is in \autoref{appendix:main}, where we show that \ref{eq:opt} can be cast as a finite-dimensional program with affine objective and constraints on the convex set \jointcav, denoted by \ref{eq:opt-jointcav}. Corollary 28.2.2 in \cite{rockafellar2015convex} implies that under \hyperlink{assumptioncs}{Assumption S} a Lagrange multiplier exists for \ref{eq:opt-jointcav}. We use this result to show that \autoref{eq:cav-lagrangian} holds. 
\subsection{Refined upper bounds}\label{sec:upper-bounds}
We now provide results which allow us to refine the upper bound in \autoref{cor:bound}, whenever a solution to \ref{eq:opt} exists. 

\subsubsection{Conflict-agreement}\label{sec:conflict-agreement}
Examples \ref{example:lobbying} and \ref{example:news} share the following feature: The objective function \objective\ and the constraint functions \vectorconstraint\ do not agree on the ranking of the distributions over posteriors. For instance, in \autoref{example:lobbying}, (partial) information revelation is optimal under the objective \objective, but it is not for the constraint function \vectorconstraint. Similarly, in \autoref{example:news}, the news platform prefers full information revelation, whereas the audience prefers partial information revelation. \autoref{prop:agreement} below shows that in the absence of such disagreement one can refine the upper bound on the cardinality of the support of the solutions to \ref{eq:opt}.

To introduce \autoref{prop:agreement}, we first introduce a notion of \emph{agreement} between the objective function \objective\ and the constraint functions, \vectorconstraint:
\begin{definition}[Agreement]\label{definition:agreement}
The tuple $(\objective,\vectorconstraint)$ is in agreement if the following holds. For any two posterior distributions $\bsplit,\bsplit^\prime\in\Delta_{\prior}(\Posteriors)$, if \objective\ prefers \bsplit\ to $\bsplit^\prime$, that is,
\[\mathbb{E}_\tau\left[\objective\right]\geq\mathbb{E}_{\bsplit^\prime}\left[\objective\right],\]
then
\begin{enumerate}
\item\label{itm:inequality} $\vectorconstraint_\inequality$ prefers \bsplit\ to $\bsplit^\prime$, that is, for all $i\in\{1,\dots,\inequality\}$, $\mathbb{E}_\bsplit\left[\const_i\right]\geq\mathbb{E}_{\bsplit^\prime}\left[\const_i\right]$, and
\item\label{itm:equality} $\vectorconstraint_\equality$ is indifferent between \bsplit\ and $\bsplit^\prime$, that is, for all $i\in\{\inequality+1,\dots,\inequality+\equality\}$, $\mathbb{E}_{\bsplit}\left[\const_i\right]=\mathbb{E}_{\bsplit^\prime}\left[\const_i\right]$.
\end{enumerate}
\end{definition}
To understand why \autoref{definition:agreement} distinguishes between inequality and equality constraints, recall that one can always express \ref{eq:opt} as an optimization problem subject to only inequality constraints. Indeed, one may replace the equality constraints, $\mathbb{E}_\bsplit[\vectorconstraint_\equality]=\side_\equality$, by two inequality constraints, $\mathbb{E}_\bsplit[\vectorconstraint_\equality]\geq\side_\equality$ and $\mathbb{E}_\bsplit[-\vectorconstraint_\equality]\geq-\side_\equality$. Under this alternative representation of \ref{eq:opt}, the requirement that equality constraints are indifferent between \bsplit\ and $\bsplit^\prime$ in \autoref{itm:equality} in \autoref{definition:agreement} follows from applying \autoref{itm:inequality} in \autoref{definition:agreement} to $\vectorconstraint_\equality$ and $-\vectorconstraint_\equality$.

A tuple (\objective,\vectorconstraint) satisfies \autoref{definition:agreement} if, for instance, (i) \objective\ and $\vectorconstraint_\inequality$ are concave, while $\vectorconstraint_\equality$ is affine, or (ii) \objective\ and $\vectorconstraint_\equality$ are convex, while $\vectorconstraint_\equality$ is affine.

\begin{prop}[Agreement leads to posterior distributions with at most $N$ beliefs]\label{prop:agreement}
Suppose (i) the tuple (\objective,\vectorconstraint) is in agreement, (ii) \hyperlink{assumptionf}{Assumption \objective} holds, and (iii) $(\prior,\side)\in\feasible$. Then, a solution $\bsplitopt$ to \ref{eq:opt} exists and satisfies that $|\supp\;\bsplitopt|\leq N$.
\end{prop}
The proof of \autoref{prop:agreement} shows that when (\objective,\vectorconstraint) are in agreement, then the distribution over posteriors that attains \cav\;\objective(\prior) (i.e., the \emph{unconstrained} solution) is feasible for \ref{eq:opt}. Thus, \autoref{definition:agreement} provides an economically principled way of understanding in which environments the constraints do not bind.

An immediate corollary of \autoref{prop:agreement} is the following. Suppose that we expand the optimization problem in \ref{eq:opt} by adding $(\vectorconstraintb,\sideb)$ such that (i) $(\objective,\vectorconstraintb)$ are in agreement and (ii) a solution exists to the new optimization problem. Then, a solution \bsplit\ exists with $|\support\;\bsplit|\leq N+\inequality+\equality$. That is, the support of the optimal posterior distribution does not increase when we add constraints that are in agreement with the objective and do not make the program infeasible.\footnote{To see this, let \bsplitopt\ denote the solution to \ref{eq:opt} and \bsplitb\ denote the solution to the program with $(\objective,\vectorconstraint,\vectorconstraintb)$ at (\prior,\side,\sideb), which exists by assumption. This means that \bsplitb\ is feasible for \ref{eq:opt}. Thus,  $\mathbb{E}_{\bsplitopt}\objective\geq\mathbb{E}_{\bsplitb}\objective$. If the latter inequality is strict, \autoref{definition:agreement} implies that \bsplitopt\ is feasible for $(\objective,\vectorconstraint,\vectorconstraintb)$ at (\prior,\side,\sideb).}

\subsubsection{Generalized Information design}\label{sec:gid}
While \autoref{prop:agreement} provides a form of agreement under which the support of the solutions to \ref{eq:opt} is at most the cardinality of the set of states, as we illustrate next, not every form of disagreement leads to adding beliefs in the support of a solution to \ref{eq:opt}:

\begin{example}[Prosecutor and judge with an outside option]\label{example:oo}
Consider the following version of the prosecutor example in \cite{kamenica2011bayesian}. Everything is as in \cite{kamenica2011bayesian} except that the judge has access to outside information. However, the judge has limited time and can either listen to the prosecution or to their own information source. We model this as the prosecutor facing a constraint: the judge has to receive at least the payoff they can achieve by using their own information source.

Formally, let  $\States=A=\{0,1\}$ denote the set of states and judge's actions. Payoffs are $u(a,\state)=\mathbbm{1}[a=\state]$ and $v(a,\state)=\mathbbm{1}[a=1]$ for the judge and the prosecutor, respectively. Let $\prior\in[0,1]$ denote the prior probability that $\state=1$. Assume that $\prior<1/2$. The judge has access to another experiment, $\bsplit^J\in\Delta_{\prior}(\Posteriors)$. Let $a^*(\belief)$ denote the judge's optimal action choice when the posterior belief is $\belief$, breaking ties if necessarily in favor of the prosecutor. Finally, let  $\hat{v}(\belief)=\sum_{\state\in\States}\belief(\state)v(a^*(\belief),\state)$ and $\hat{u}(\belief)=\mathbb{E}_{\belief}\left[u(a^*(\belief),\cdot)\right]$ denote the prosecutor and judge's indirect utility functions. Then, the prosecutor's optimal payoff follows from solving the following problem:\label{page-oo-typo}
\begin{align}\label{eq:opt-oo}\tag{\OPT$_{KG}$}
&\max_{\tau\in\Delta_{\prior}\Posteriors}\mathbbm{E}_\tau\left[\hat{v}(\belief)\right]\\
\text{ s.t. }&\mathbb{E}_\tau\left[\hat{u}(\belief)\right]\geq\mathbb{E}_{\tau^J}\left[\hat{u}(\belief)\right].\nonumber
\end{align}
Since the prosecutor can design any experiment, the prosecutor can always replicate the judge's information source. Thus, without loss of generality, the judge uses the prosecutor's experiment, while the prosecutor offers the judge an experiment that is at least as valuable to the judge as their own information source.

Appealing to \autoref{prop:lagrangian}, we setup the Lagrangian objective $\hat{v}(\belief)+\dual\hat{u}(\belief)$, which we depict in \autoref{fig:lagrangian-oo}.
\begin{figure}[th!]
\centering
\subfloat[Objective and constraint functions]{
\begin{tikzpicture}[scale=0.8]
\begin{axis}[axis lines=middle,xmin=-0.05,xmax=1.1,ymin=-0.05,ymax=1.05,xlabel=$\belief$,xtick={0,0.5,1},ytick={0,0.5,1},x label style={at={(axis description cs:1,-0.05)}},
    y label style={at={(axis description cs:-0.05,1)}},
    width=8cm,
        height=8cm, legend style={fill=none, at={(0.8,0.2)},anchor=west}]
\addplot[color=blue,thick,domain=0:0.5]{0};
\addlegendentry{$\hat{v}$};
\addplot[color=red,thick,domain=0.5:1]{x};
\addlegendentry{$\hat{u}$};
\addplot[color=blue,thick,mark=o] coordinates {(0.5,0)};
\addplot[color=blue,thick,mark=*] coordinates {(0.5,1)};
\addplot[color=blue,thick,domain=0.5:1]{1};
\addplot[color=red,thick,domain=0:0.5]{1-x};
\end{axis}
\end{tikzpicture}\label{fig:oo-f-g}
}
\subfloat[$\dual=0.5$]{\begin{tikzpicture}[scale=0.8]
\begin{axis}[axis lines=middle,xmin=-0.05,xmax=1.1,ymin=-0.05,ymax=3.1,xlabel=$\belief$, ytick={},xtick={0,0.5,1},ytick={0.5,1,2},ylabel=$\hat{v}+0.5\hat{u}$,x label style={at={(axis description cs:1,-0.05)}},
    y label style={at={(axis description cs:-0.25,1)}},
    width=8cm,
        height=8cm]
\addplot[color=blue,domain=0:0.5]{0.5*(1-x)};
\addplot[color=blue,thick,mark=o] coordinates {(0.5,0.25)};
\addplot[color=blue,thick,mark=*] coordinates {(0.5,1.25)};
\addplot[color=blue,domain=0.5:1]{1+0.5*x};
\addplot[color=red,dashed,thick] coordinates {(0,0.5) (0.5,1.25)};
\end{axis}
\end{tikzpicture}\label{fig:oo-low-dual}}

\subfloat[$\dual=1=\dual\conjugate$]{
\begin{tikzpicture}[scale=0.8]
\begin{axis}[axis lines=middle,xmin=-0.05,xmax=1.1,ymin=-0.05,ymax=3.1,xlabel=$\belief$, ytick={},xtick={0,0.5,1},ytick={0.5,1,2},ylabel=$\hat{v}+\hat{u}$,x label style={at={(axis description cs:1,-0.05)}},
    y label style={at={(axis description cs:-0.25,1)}},
    width=8cm,
        height=8cm]
\addplot[color=blue,thick,domain=0:0.5]{1-x};
\addplot[color=blue,thick,mark=o] coordinates {(0.5,0.5)};
\addplot[color=blue,thick,mark=*] coordinates {(0.5,1.5)};
\addplot[color=blue,thick,domain=0.5:1]{1+x};
\addplot[color=red,dashed,thick] coordinates {(0,1) (0.5,1.5)};
\end{axis}
\end{tikzpicture}\label{fig:oo-opt-dual}
}
\subfloat[$\dual=2$]{
\begin{tikzpicture}[scale=0.8]
\begin{axis}[axis lines=middle,xmin=-0.05,xmax=1.1,ymin=-0.05,ymax=3.1,xlabel=$\belief$,ytick={},xtick={0,0.5,1}, ytick={0.5,1,2},ylabel=$\hat{v}+2\hat{u}$,x label style={at={(axis description cs:1,-0.05)}},
    y label style={at={(axis description cs:-0.3,1)}},
    width=8cm,
        height=8cm]
\addplot[color=blue,thick,domain=0:0.5]{2*(1-x)};
\addplot[color=blue,thick,mark=o] coordinates {(0.5,1)};
\addplot[color=blue,thick,mark=*] coordinates {(0.5,2)};
\addplot[color=blue,thick,domain=0.5:1]{1+2*x};
\addplot[color=red,dashed,thick] coordinates {(0,2) (1,3)};
\end{axis}
\end{tikzpicture}
\label{fig:oo-high-dual}}
\caption{Lagrangian in \autoref{example:oo}; the optimal multiplier is $\dual\conjugate=1$.}\label{fig:lagrangian-oo}
\end{figure}
Relying once again on \autoref{observation:visual}, it is possible to show that a unique Lagrange multiplier \dual\conjugate\ exists such that the solution may involve more than two posterior beliefs: This corresponds to $\dual\conjugate=1$ and is depicted in \autoref{fig:oo-opt-dual}. In \autoref{fig:oo-opt-dual}, infinitely many beliefs are candidates for the support of an optimal solution, namely, $\{0\}\cup[1/2,1]$. Note, however, that $\hat{v}$ and $\hat{u}$ are linear on $[1/2,1]$ (\autoref{fig:oo-f-g}), so that there is no loss of optimality or feasibility in inducing at most one posterior belief in $[1/2,1]$, which will be chosen to satisfy the judge's participation constraint. It follows that a solution to \ref{eq:opt-oo} exists with at most binary support.
\end{example}

\autoref{definition:gid} distills the mathematical property that lies behind \autoref{example:oo}:
\begin{definition}[Generalized information design environment]\label{definition:gid}
A \emph{$K$-generalized information design environment} is a tuple $(\objective,\vectorconstraint_\inequality,\vectorconstraint_\equality)$ such that a partition $(\Delta_k)_{k=1}^K$ of \Posteriors\ exists satisfying the following properties:
\begin{enumerate}
\item For each $k\in\{1,\dots,K\}$, $\Delta_k$ is convex,
\item For each $k\in\{1,\dots, K\}$, $(\objective,\vectorconstraint_\inequality)$ is concave on $\Delta_k$,
\item For each $k\in\{1,\dots, K\}$, $\vectorconstraint_\equality$ is affine on $\Delta_k$.
\end{enumerate}
\end{definition}
\autoref{example:oo} is an instance of a $2$-generalized information design environment for the partition $\Delta_1=[0,1/2)$ and $\Delta_2=[1/2,1]$. Indeed, both the objective function, $\hat{v}$, and the constraint function, $\hat{u}$, are linear on each element of the partition.\footnote{\autoref{example:lobbying} is a $4$-generalized information design environment, where $\Delta_1=[0,\nicefrac{1}{3}),\Delta_2=[\nicefrac{1}{3},\nicefrac{2}{3}),\Delta_3=\{\nicefrac{2}{3}\},\Delta_4=(\nicefrac{2}{3},1]$.\label{ftn:gid-examples}}

To gain some intuition behind \autoref{definition:gid}, consider a standard Bayesian persuasion problem, where a receiver possesses $K$ actions, $\{a_1,\dots,a_K\}$. Then, $\Delta_k$ represents the set of beliefs for which $a_k$ is optimal for the receiver, so that $\Delta_k$ is a convex set and $\cup_{k=1}^K\Delta_k=\Posteriors$. While $\left(\Delta_k\right)_{k=1}^K$ is not necessarily a partition of \Posteriors\ (e.g., the receiver may have multiple best responses at a given belief), in applications we usually work with a selection out of the receiver's best response correspondence, $a\conjugate(\belief)$, which we can use to redefine  the collection $(\Delta_k)_{k=1}^K$ to be a partition.

Suppose now that the tuple $(\objective,\vectorconstraint)$ are indirect utilities. That is, there exist $(\objective^\prime,\vectorconstraint^\prime):A\times\States\mapsto\mathbb{R}$, such that
\begin{align*}
\objective(\belief)=\sum_{\state\in\States}\belief(\state) \objective^\prime(a\conjugate(\belief),\state),\vectorconstraint(\belief)=\sum_{\state\in\States}\belief(\state)\vectorconstraint^\prime(a\conjugate(\belief),\state).
\end{align*}
Then, it is immediate to see that on $\Delta_k$ -- that is, holding the receiver's best response fixed,-- $(\objective,\vectorconstraint)$ are linear functions of \belief. The term \emph{generalized information design} references that we allow the objective, \objective, and the inequality constraints, $\vectorconstraint_\inequality$, to be concave, rather than just linear.

In standard Bayesian persuasion, one can usually appeal to the revelation principle \citep{myerson1982optimal} to bound the number of posteriors by the cardinality of the action set. \autoref{prop:gid} shows that this insight extends to generalized information design environments, providing us with an additional bound on cardinality of the support of the solutions to \ref{eq:opt} (see \autoref{cor:ub-gid}).
\begin{prop}[Generalized information design]\label{prop:gid}
Assume $(\objective,\vectorconstraint_\inequality,\vectorconstraint_\equality)$ is a $K-$generalized information design environment.  If a solution to \ref{eq:opt} exists, then a solution $\bsplitopt$ to \ref{eq:opt} exists with $|\supp\;\bsplitopt|\leq K$.
\end{prop}
Together with \autoref{cor:bound}, \autoref{prop:gid} implies the following:
\begin{cor}[Support upper-bound in Generalized Information Design problems]\label{cor:ub-gid}
In a $K-$generalized information design environment, if a solution to \ref{eq:opt} exists, then a solution $\bsplitopt$ to \ref{eq:opt} exists with $|\supp\;\bsplitopt|\leq\min\{ K,N+\inequality+\equality\}$.
\end{cor}
\autoref{example:sm} in \autoref{appendix:examples} illustrates that in $K$-generalized information design environments with $N<K$, optimal policies can induce more posteriors than the number of states, but less than the number of states plus (binding) constraints (that is, $N<K<N+\inequality+\equality$). 

\section{Applications}\label{sec:applications}
\autoref{sec:applications} illustrates how the tools in \autoref{sec:main} can be brought to bear in problems that seemingly do not fit the statement of \ref{eq:opt}, but by reformulating them as constrained information design problems, their analysis can be simplified. \autoref{sec:limited-commitment} studies a simple instance of mechanism design with limited commitment as in \cite{bester2007contracting}. We rely on the revelation principle in \cite{doval2020mechanism} to formulate a program that combines elements of information design and mechanism design and characterizes the principal's optimal mechanism. Armed with \autoref{theorem:tomalalemma} and its corollaries, we bound the the cardinality of the support of the principal's optimal mechanism. Instead, \autoref{sec:informed-receiver} studies Bayesian persuasion of a privately informed receiver. While the receiver's incentive constraints do not a priori fit those in \ref{eq:opt}, \autoref{prop:informed-receiver} illustrates how the results so far can be used to bound the cardinality of the support of each experiment in the \emph{menu} offered by the sender.
%
%
%

\subsection{Mechanism design with limited commitment}\label{sec:limited-commitment}


\autoref{sec:limited-commitment} showcases how \autoref{theorem:tomalalemma} can be leveraged to inform the characterization of optimal mechanisms in settings in which the principal has limited commitment. To keep the presentation simple and to facilitate the comparison with other work in the literature, we present the results in the context of a reduced-form representation of limited commitment based on the model in \cite{bester2007contracting}. 

Consider the problem of a principal who interacts with a privately informed agent, who knows the state of the world. Let $\prior\in\Posteriors$ denote the principal's prior belief about the state of the world. The interaction lasts for two periods, $t\in\{1,2\}$. In each period $t$, as a result of the interaction, an allocation $\allocation_t\in\Allocations_t$ is determined, where $\Allocations_t$ is the set of allocations in period $t$. There is a correspondence 
$\constraint:\Allocations_1\rightrightarrows\Allocations_2$ that describes the set of feasible period $2$ allocations as a function of the allocation in period $1$. Below, the allocation $\allocation_2\in\Allocations_2$ captures in reduced form the principal's limited commitment. Let $v(\allocation_1,\allocation_2,\state)$ and $u(\allocation_1,\allocation_2,\state)$ denote the principal and the agent's payoff, respectively, when the allocation is $(\allocation_1,\allocation_2)$ and the state of the world is \state. We assume an allocation $(\allocation_1^*,\allocation_2^*)$  exists such that $u(\allocation_1^*,\allocation_2^*,\state)=0$ for all $\state\in\States$. This allocation plays the role of the outside option in what follows.\footnote{Throughout, we make the following technical assumptions. First, the set of allocations $\Allocations_1,\Allocations_2$ are compact Polish spaces. Second, endowing product sets with their product $\sigma$-algebra, we assume that the principal and the agent's utility functions are bounded measurable functions. We assume that the principal's utility is continuous in $\allocation_2$ for each $(\allocation_1,\state)$.  Third, the correspondence \constraint\ is measurable and for each $\allocation_1\in\Allocations_1$, $\constraint(\allocation_1)$ is compact. Finally, for any two measurable spaces $X$ and $Y$ , a mapping $\zeta:X\mapsto\Delta(Y)$ is a \emph{transition probability} from $X$ to $Y$ if, for any measurable $C\subseteq Y$, $\zeta(C|x)\equiv\zeta(x)(C)$ is a measurable real valued function of $x\in X$.\label{ftn:technical}}
\paragraph{Mechanisms} A mechanism \mechanism\ consists of a set of input messages $\inputt$, a set of output messages $\outputt$, and a device $\joint:\inputt\mapsto\Delta(\outputt\times\Allocations_1)$, which associates to each input message $m\in\inputt$ a distribution over output messages and allocations. 

\paragraph{Timing} The game proceeds as follows: In period $1$, after the agent observes the state $\state\in\States$, the principal offers the agent a mechanism, \mechanism. After observing the mechanism, the agent decides whether to accept or reject. If she rejects the mechanism, an allocation $(\allocation_1^*,\allocation_2^*)\in\Allocations_1\times\Allocations_2$ is implemented. If instead she accepts the mechanism, she \emph{privately} submits an input message to the mechanism. This message determines the distribution $\joint(\cdot|m)$ from which the output message and the allocation are drawn. In period $2$, after observing the output message and the allocation, the principal chooses an allocation $\allocation_2\in\Allocations_2$.

Our objective is to characterize the optimal mechanism for the principal under the solution concept of Perfect Bayesian equilibrium. In particular, the principal's choice of the allocation in period $2$ must be sequentially rational.  Indeed, for each $\allocation_1\in\Allocations_1$, denote by
\begin{align}\label{eq:best-response}
\allocation_2(\allocation_1,\belief)\in\bestresponse_2(\allocation_1,\posterior)\equiv\arg\max_{\allocation_2\in\constraint(\allocation_1)}\sum_{\state\in\States}\belief(\state)v(\allocation_1,\allocation_2,\state),
\end{align}
 a solution to the principal's problem in period $2$ when his belief about the state of the world is \belief. The assumptions in \autoref{ftn:technical} imply the above problem is well-defined. In a slight abuse of notation, let $\bestresponse_2$ denote the set of all selections from the  best response correspondence $\bestresponse_2(\cdot)$. 

\begin{remark}[Mechanism design with downstream interactions]\label{rem:downstream}
Whereas below we emphasize the limited commitment interpretation, by replacing the payoff function $v(\cdot)$ with another function $w(\cdot)$ in \autoref{eq:best-response}, this model also captures in reduced form settings where an upstream mechanism designer chooses a mechanism anticipating that a downstream third party observes the mechanism's allocation and takes an action $\allocation_2\in\Allocations_2$ in response. The downstream third party may represent another principal as in \cite{calzolari2006optimality,pavan2009sequential}, or an aftermarket as in \cite{calzolari2006monopoly,dworczak2020mechanism}. 
\end{remark}

\paragraph{Revelation principle} Theorem 1 in \cite{doval2020mechanism} implies the principal-optimal Perfect Bayesian equilibrium can be characterized as the solution to a constrained optimization program (see \ref{eq:opt-limited-commitment} below). Indeed, Theorem 1 in \cite{doval2020mechanism} has the following three implications. First it is without loss of generality to restrict attention to \emph{direct Blackwell mechanisms}, where the set of input and output messages are the set of states and posterior beliefs, respectively, i.e., $\inputt=\States$ and $\outputt=\Posteriors$. Furthermore, the device \joint\ can be decomposed into two transition probabilities, a disclosure rule, $\beta:\States\mapsto\Delta\Posteriors$, and an allocation rule, $\alpha:\Posteriors\mapsto\Delta(\Allocations_1)$. 
Second, it is without loss of generality to restrict attention to equilibrium strategies such that the agent finds it is optimal to participate and truthfully report the state of the world. Finally, when the mechanism outputs a belief \belief, this is the belief that would result from Bayes' rule when the principal observes output message \belief, and the agent participates and truthfully reports her type. When for each \state, the distribution $\beta(\cdot|\state)$ has finite support\label{page-beta-finite}, this is equivalent to requiring that
\begin{align}\label{eq:post-lc}
\belief(\state)=\frac{\prior(\state)\beta(\belief|\state)}{\sum_{\stateb\in\States}\prior(\stateb)\beta(\belief|\stateb)}.
\end{align}

We can write the principal's problem as follows:\footnote{\cite{bester2007contracting} analyze a version of the problem \ref{eq:opt-limited-commitment} with the following differences. First, instead of letting the set of output messages be the set of beliefs, they leave the set $S$ unspecified. Therefore, they only analyze the principal's optimal mechanism within those that use signals in $S$. Second, because the set $S$ is unspecified, their program has an additional constraint: the choice of $\allocation_2$ has to be optimal given the realization of $s$ and the period-$1$ allocation.  Finally, they do not allow for randomized allocations. \cite{doval2020mechanism} show that this may be with loss of generality (\citealp{strausz2003deterministic} also shows the importance of allowing for randomization for the standard version of the revelation principle to hold). }
\small
\begin{align}\label{eq:opt-limited-commitment}\tag{$\mathcal{P}$}
&\max_{\beta:\Types\mapsto\Delta\Posteriors,\alpha:\Posteriors\mapsto\Delta(\Allocations_1),\allocation_2\in\bestresponse_2}\sum_{\state\in\States}\prior(\state)\mathbb{E}_{\beta(\cdot|\state)}\left[\mathbb{E}_{\alpha(\cdot|\belief)}\left[v(\allocation_1,\allocation_2(\allocation_1,\belief),\state)\right]\right]\\
&\text{s.t.}\left\{\begin{array}{ll}
(\forall\state\in\States)&\mathbb{E}_{\beta(\cdot|\state)}\left[\mathbb{E}_{\alpha(\cdot|\belief)}\left[u(\allocation_1,\allocation_2(\allocation_1,\belief),\state)\right]\right]\geq 0\\
(\forall\state\in\States)(\forall\stateb\neq\state)&\mathbb{E}_{\beta(\cdot|\state)-\beta(\cdot|\stateb)}\left[\mathbb{E}_{\alpha(\cdot|\belief)}\left[u(\allocation_1,\allocation_2(\allocation_1,\belief),\state)\right]\right]\geq0
\end{array}\right.,\nonumber
\end{align}
\normalsize
where the two sets of constraints are the agent's participation and truthtelling constraints. Furthermore, the transition probability $\beta$ must satisfy that for all measurable subsets \measurablem\ of \Posteriors\ and all subsets \measurablet\ of \States, 
\begin{align*}
\sum_{\stateb\in\measurablet}\beta(\measurablem|\stateb)\prior(\stateb)=\sum_{\state\in\States}\int\posterior(\measurablet)\beta(d\posterior|\state)\prior(\state).
\end{align*}

\paragraph{Limited commitment as constrained information design.}To show how \autoref{theorem:tomalalemma} can inform the solution to \ref{eq:opt-limited-commitment} we first show how to write the principal's optimization problem as one in which he chooses a Bayes' plausible distribution over posteriors and an allocation rule $\alpha:\Posteriors\mapsto\Delta(\Allocations_1)$. For any measurable subset \measurablem\ of \Posteriors, and for any subset \measurablet\ of \States, let $\jointp\in\Delta(\States\times\Posteriors)$ denote the following measure:
\begin{align*}
\jointp(\measurablet\times\measurablem)=\sum_{\state\in\measurablet}\beta(\measurablem|\state)\prior(\state).
\end{align*}
The disintegration theorem (Proposition 3.6 in \citealp{crauel2002random}) implies a distribution $\tau\in\Delta\Posteriors$ exists such that
\begin{align*}
\jointp(\measurablet\times\measurablem)=\int_{\measurablet}\left(\sum_{\state\in\measurablet}\belief(\state)\right)\tau(d\belief).
\end{align*}
It follows that for all $\state\in\States$ and all measurable subsets \measurablem\ of \Posteriors, we have
\begin{align*}
\beta(\measurablem|\state)\prior(\state)=\int_{\measurablet}\belief(\state)\tau(d\belief),
\end{align*}
and we can write the agent's payoff when the state is \state\ and she reports \stateb\ as follows:
\begin{align*}
\mathbb{E}_{\beta(\cdot|\stateb)}\left[\mathbb{E}_{\alpha(\cdot|\belief)}\left[u(\allocation_1,\allocation_2(\allocation_1,\belief),\state)\right]\right]=\mathbb{E}_{\tau}\left[\mathbb{E}_{\alpha(\cdot|\belief)}\left[\frac{\belief(\stateb)}{\prior(\stateb)}u(\allocation_1,\allocation_2(\allocation_1,\posterior),\state)\right]\right].
\end{align*}
These steps allow us to express the principal's objective and the agent's incentive and participation constraints as expectations under the same measure, $\bsplit\in\Delta_{\prior}(\Posteriors)$. Indeed,
we can write the principal's problem similar to the problem in \ref{eq:opt}:
\begin{align}\label{eq:opt-limited-commitment-beliefs}\tag{\OPT$_{LC}$}
&\max_{\bsplit\in\Delta_{\prior}\Posteriors,\alpha:\Posteriors\mapsto\Delta(\Allocations_1),\allocation_2\in\bestresponse_2}\mathbb{E}_\tau\left[\mathbb{E}_{\alpha(\cdot|\belief)}\left[\sum_{\state\in\States}\belief(\state)v(\allocation_1,\allocation_2(\allocation_1,\belief),\state)\right]\right]\\
&s.t.\left\{
\begin{array}{ll}
(\forall\state\in\States)&\mathbb{E}_{\tau(\cdot)}\left[\mathbb{E}_{\alpha(\cdot|\belief)}\left[\frac{\belief(\state)}{\prior(\state)}u(\allocation_1,\allocation_2(\allocation_1,\belief),\state)\right]\right]\geq 0\\
(\forall\state\in\States)(\forall\stateb\neq\state)&\mathbb{E}_{\tau(\cdot)}\left[\mathbb{E}_{\alpha(\cdot|\belief)}\left[\left(\frac{\belief(\state)}{\prior(\state)}-\frac{\belief(\stateb)}{\prior(\stateb)}\right)u(\allocation_1,\allocation_2(\allocation_1,\belief),\state)\right]\right]\geq0\end{array}\right.\nonumber.
\end{align}


The resulting program \ref{eq:opt-limited-commitment-beliefs} allows us to highlight the connection between mechanism design with limited commitment and the literature on information design. After all, the designer can be thought of as a sender who designs an information structure for a receiver, who happens to be his future self.  However, there are differences.  In our setting, the principal (the sender in \citealp{kamenica2011bayesian}) also chooses a distribution of period-1 allocations for each posterior he induces.  In addition, the principal cannot implement any Bayes' plausible distribution over posteriors, but only those that satisfy the incentive compatibility and participation constraints of the agent, which, in turn, depend on the resulting period-2 allocation.

In what follows, we focus on the case in which the agent's preferences satisfy a version of \emph{increasing differences}, which takes into account that the agent faces lotteries over allocations: 
\begin{definition}[\citealp{bester2007contracting,celik2015implementation,kartik2017single}]\label{definition:med}
The family $\{u(\cdot,\state):\state\in\States\}$ satisfies \emph{monotonic expectational differences} if for any two distributions $P,Q\in\Delta(\Allocations_1\times\Allocations_2)$, $\int u(\cdot,\state_i)d(P-Q)$ is monotone in $i$.
\end{definition}
\cite{kartik2017single} show that $u$ satisfies monotonic expectational differences if, and only if, it takes the form, $u(\allocation_1,\allocation_2,\state_i)=b(\state_i)h_1(\allocation_1,\allocation_2)+h_2(\allocation_1,\allocation_2)+c(\state_i)$, where $h_1,h_2$ are finitely integrable and $b$ is monotonic. Without loss of generality, assume that $b$ is weakly increasing, so that $\state_1$ is the agent's ``lowest type.''\footnote{We also assume that $h_1(\allocation_1^*,\allocation_2^*)=\min_{(\allocation_1,\allocation_2):\allocation_2\in\constraint(\allocation_1)}h_1(\allocation_1,\allocation_2)$. This allows us to conclude that whenever the lowest type, $\state_1$ participates of the mechanism, then all types participate of the mechanism.\label{ftn:participation}}

Like increasing differences in mechanism design with commitment, monotonic expectational differences imply that the solutions to \ref{eq:opt-limited-commitment} coincide with the solutions to a much simpler program, which imposes only a subset of the incentive compatibility constraints:
\begin{prop}\label{prop:med}
If $\{u(\cdot,\state):\state\in\States\}$ satisfies monotonic expectational differences, then to characterize the solution to \ref{eq:opt-limited-commitment}, it suffices to guarantee that
\begin{enumerate}
\item The agent's participation constraint holds when the state is $\state_1$, and
\item Adjacent incentive compatibility constraints are satisfied.
\end{enumerate}
\end{prop}
See \autoref{appendix:limited-commitment} for a proof. We then obtain the following corollary:
\begin{cor}\label{cor:bound-limited-commitment-1}
Any solution to \ref{eq:opt-limited-commitment} utilizes at most $3N-1$ posteriors.
\end{cor}
\begin{remark}[Connection to the literature]\label{remark:med-literature}  \autoref{prop:med} has antecedents in the literature.  Assuming that the agent's utility function has the form characterized by \cite{kartik2017single}, \cite{bester2007contracting} show that adjacent incentive constraints imply global incentive compatibility constraints. However, they do not show that the lowest type's participation constraint implies the other participation constraints, which requires an assumption like the one we make in \autoref{ftn:participation}. Instead, \cite{celik2015implementation} assumes the condition in  \autoref{definition:med} holds and states without proof that it implies that adjacent incentive compatibility constraints imply global incentive compatibility constraints. Besides providing a complete proof of the result, \autoref{prop:med} closes the link between the two aforementioned papers via the result in \cite{kartik2017single} which shows that the only utility functions that satisfy the condition in \cite{celik2015implementation} are those assumed by \cite{bester2007contracting}. 
\end{remark}

\paragraph{Transferable utility:} Transferable utility is a common assumption in mechanism design. In what follows, we show how this assumption further simplifies the characterization of an optimal mechanism by reducing in some instances the number of posteriors that the mechanism employs. Therefore, we make the following assumptions in the remainder of this section. First, the set of period $1$ allocations is given by $\Allocations_1=\Allocations_1^\prime\times\mathbb{R}_+$, where the second coordinate denotes a payment from the agent to the principal. We denote an element of $\Allocations_1$ by $\allocation_1=(\allocationb_1,\transfer)$. Second, we assume that $\constraint((\allocationb_1,\transfer))=\constraint(\allocationb_1)$. Finally, we assume that the agent and the principal's payoffs can be written as follows:
\begin{align*}
v(\allocation_1,\allocation_2,\state)=\tilde{v}(\allocationb_1,\allocation_2,\state)+\transfer,\;
u(\allocation_1,\allocation_2,\state)=\tilde{u}(\allocationb_1,\allocation_2,\state)-\transfer.
\end{align*}
Transferable utility implies that focusing on mechanisms that do not randomize on transfers is without loss of generality. Hereafter, we replace \transfer\ with its expectation under the mechanism when the posterior is \belief, $\transfer(\belief)$, and we let $\tilde{\alpha}:\Posteriors\mapsto\Delta(\Allocations_1^\prime)$ denote the marginal of $\alpha$ over $\Allocations_1^\prime$.

The following result follows from \autoref{prop:med}:
\begin{cor}\label{cor:ir-binding}
Suppose that the family $\{u(\cdot,\state):\state\in\States\}$ satisfies monotonic expectational differences and utility is transferable. Then, the participation constraint is binding for $\state_1$.
\end{cor}

Under the assumptions of monotonic expectational differences and transferable utility, we could further simplify \ref{eq:opt-limited-commitment} by showing that \emph{downward-looking} incentive constraints always bind at the optimum. This then justifies the study of the so-called relaxed program:
\small
\begin{align}\label{eq:relaxed}\tag{$\mathcal{R}$}
&\max_{\tau\in\Delta_{\prior}\Posteriors,\tilde{\alpha}:\Posteriors\mapsto\Delta(\Allocations_1^\prime),\transfer:\Posteriors\mapsto\reals,\allocation_2(\cdot)\in\bestresponse_2(\cdot)}\mathbb{E}_{\tau}\left[\mathbb{E}_{\tilde{\alpha}(\cdot|\belief)}\left[\sum_{\state\in\States}\belief(\state)\tilde{v}(\allocationb_1,\allocation_2(\allocationb_1,\belief),\state)+\transfer(\belief)\right]\right]\\
&\text{s.t.}\left\{\begin{array}{ll}
&\mathbb{E}_{\tau}\left[\mathbb{E}_{\alpha(\cdot|\posterior)}\left[\frac{\posterior(\state_1)}{\prior(\state_1)}\left(\tilde{u}(\allocationb_1,\allocation_2(\allocationb_1,\posterior),\state)-\transfer(\posterior)\right)\right]\right]=0\\
(\forall i\in\{2,\dots,N\})&\mathbb{E}_{\tau}\left[\mathbb{E}_{\tilde{\alpha}(\cdot|\posterior)}\left[\left(\frac{\posterior(\state_i)}{\prior(\state_i)}-\frac{\posterior(\state_{i-1})}{\prior(\state_{i-1})}\right)\left(\tilde{u}(\allocationb_1,\allocation_2(\allocationb_1,\posterior),\state_i)-\transfer(\posterior)\right)\right]\right]=0
\end{array}\right.,\nonumber
\end{align}
\normalsize
which is obtained by dropping the \emph{monotonicity constraints}:\footnote{The constraints in \autoref{eq:monotonicity} are obtained by combining the restriction that $\state_i$ does not want to report $\state_{i-1}$ and $\state_{i-1}$ does not want to report $\state_i$. Under \autoref{definition:med}, the binding downward-looking incentive constraints together with the monotonicity constraints imply the local constraints in \autoref{prop:med}.\label{ftn:monfnt}}\small
\begin{align}\label{eq:monotonicity}\tag{M}
\mathbb{E}_{\tau}\left[\mathbb{E}_{\alpha(\cdot|\posterior)}\left[\left(\frac{\posterior(\state_i)}{\prior(\state_i)}-\frac{\posterior(\state_{i-1})}{\prior(\state_{i-1})}\right)\left(\tilde{u}(\allocationb_1,\allocation_2(\allocationb_1,\posterior),\state_i)-\tilde{u}(\allocationb_1,\allocation_2(\allocationb_1,\posterior),\state_{i-1})\right)\right]\right]\geq0,
\end{align}\normalsize
for each $i\in\{2,\dots, N\}$. We can use the binding constraints to substitute the transfers out of the principal's program and obtain the following:
\begin{prop}\label{prop:vs}
The solution to the relaxed program uses at most $N$ posteriors. 
\end{prop}
The proof of \autoref{prop:vs} is in \autoref{appendix:limited-commitment}. It follows from two observations. First, once we substitute the transfers out of the principal's payoff, we are left with an expression that only depends on the distribution over posteriors induced by the mechanism and the portion of the allocation rule that corresponds to $\Allocations_1^\prime$. Second, since we are ignoring the monotonicity constraints, one can solve for the optimal $\alpha$ by pointwise maximization since direct Blackwell mechanisms separate the design of the mechanism's information structure (represented by \bsplit) from the design of the allocation rule (represented by $\alpha$). We are then left with a function that depends only on the distribution over posterior beliefs, that is, a standard Bayesian persuasion problem. The proof of \autoref{prop:vs} also suggests how the principal in period 1 chooses $\allocation_2$ when the principal is indifferent in period 2: ties are broken in favor of maximizing the virtual surplus.

\begin{remark}[Relaxed program and limited commitment] \autoref{prop:relaxed-more-conditions} in \autoref{appendix:limited-commitment} provides a necessary and sufficient condition under which the solution to the relaxed program can be implemented whenever it satisfies the monotonicity constraints. Whereas in mechanism design with commitment, transfers exist that implement the solution to the relaxed program as long as it satisfies the monotonicity constraints, \ref{eq:monotonicity}, this is not necessarily the case under limited commitment, as we illustrate in an example in the supplementary material.\footnote{The example in \cite{supplement} provides a counterexample to the claim in \cite{bester2007contracting} that whenever it satisfied the monotonicity constraints, the solution to the relaxed program can be implemented as a solution to \ref{eq:opt-limited-commitment}.} To see this, note that in the relaxed program the binding downward-looking incentive constraints together with $\state_1$'s participation constraint impose $N$ restrictions on the transfers $\{\transfer(\posterior):\posterior\in\Posteriors\}$. However, the solution to the relaxed program might use less than $N$ posteriors. Therefore, finding transfers $\transfer(\posterior)$ that satisfy all constraints may not possible.\footnote{This is never an issue in mechanism design with commitment: Without loss of generality, we can always have one transfer for each type.} Alternatively, not all downward-looking constraints may bind in the optimal mechanism.\footnote{Fortunately, the above is not an issue when there are two types or a continuum of types. In both cases, it is possible to show that downward looking constraints bind (see \citealp{doval2020mechanism}).}
\end{remark}

In many instances, however, the solution to \ref{eq:relaxed} will fail to satisfy the monotonicity constraints, \ref{eq:monotonicity}. As we show next, it is enough to add as many posteriors as binding monotonicity constraints at the optimum:\label{page-nec-suff}
\begin{prop}\label{prop:postmon}
Consider the program obtained by adding the monotonicity constraints \ref{eq:monotonicity} to the relaxed program \ref{eq:relaxed}. The solution to the new program uses at most $N+\binding$ posteriors, where $\binding$ is the number of binding constraints at the optimum.
\end{prop}
\autoref{prop:postmon} follows immediately  from \autoref{cor:bound} and \autoref{cor:tomalalemmadrop}.

\subsection{Persuasion of a privately informed receiver}\label{sec:informed-receiver}
Consider an information designer who controls the release of information about a state of the world $\state\in\States$ and faces a privately informed agent. Let $\Theta$ denote a finite set of agent types and let $q_\theta$ denote the probability that the agent is of type $\theta$. Let $M=|\Theta|$. As in \autoref{sec:limited-commitment}, let \prior\ denote the prior belief over \States. We assume that the state of the world \state\ and the agent's type \type\ are independently distributed. 

While the designer controls the release of information about the state of the world, the agent is the one who ultimately takes actions. That is, after observing the information released by the designer, the agent selects an action \action\ from a compact set $A$. Let $u(a,\theta,\state)$ and $v(a,\theta,\state)$ denote the agent and the designer's payoffs, respectively, when the agent takes action \action, the agent's type is \type, and the state of the world is \state. We assume that both functions are continuous in \action\ for each $(\type,\state)\in\Types\times\States$.

For each $\theta\in\Theta$, let
\begin{align*}
a^*(\belief,\theta)\in\arg\max_{a\in A}\sum_{\state\in\States}\belief(\state)u(a,\theta,\state),
\end{align*}
denote the agent's optimal action choice when her type is $\theta$ and her belief about $\state$ is given by $\belief$. Let
\begin{align*}
U(\belief,\type)=\sum_{\state\in\States}\belief(\state)u(\action^*(\belief,\type),\type,\state),
\end{align*}
denote the agent's optimal payoff when her type is \type\ and her belief about \state\ is given by \belief. Whenever necessary, we assume that the agent breaks ties in favor of the designer.

%

The information designer designs a menu of experiments $\tau: \Types \to \Delta_{\prior}\Posteriors$ to solve:\footnote{\autoref{lemma:rp} in \autoref{appendix:informed-receiver} shows that it is without loss of generality to focus on experiments where the set of signals is the space of beliefs over \States.}
\begin{align}\label{eq:opt-informed-receiver}
\max_{\tau: \Types \to \Delta_{\prior}\Posteriors}&\sum_{\type\in\Types}q_\type\sum_{\belief\in\Posteriors}\tau(\belief,\type)\sum_{\state\in\States}\belief(\state)v(\action^*(\belief,\type),\type,\state)\\
&\text{ s.t. }
(\forall\type\in\Types)(\forall\typeb\neq\type)\mathbb{E}_{\tau(\type,\cdot)}[U(\belief,\type)]\geq\mathbb{E}_{\tau(\typeb,\cdot)}[U(\belief,\type)].\nonumber
\end{align}
That is, the designer chooses an experiment to maximize his payoff subject to two constraints. First, for each type $\type\in\Types$, the experiment must induce a Bayes' plausible distribution over posteriors. Second, each type $\type\in\Types$ must prefer their experiment over the one offered to types \typeb\ other than \type.

\autoref{prop:informed-receiver} illustrates how \autoref{theorem:tomalalemma} can be used to simplify the solution to the problem in \autoref{eq:opt-informed-receiver}:
\begin{prop}\label{prop:informed-receiver}
The designer's optimal payoff can be found from the solution to
\begin{align}\label{eq:optbis}\tag{\OPT$_{IR}$}
&\max_{\{u_\type\}_{\type\in\Types}}\max_{\tau: \Types \to \Delta_{\prior}\Posteriors}\sum_{\type\in\Types}q_\type\sum_{\belief\in\Posteriors}\tau(\belief,\type)\sum_{\state\in\States}\belief(\state)v(\action^*(\belief,\type),\type,\state)\\
&\text{ s.t. }\left\{\begin{array}{ll}(\forall\type\in\Types)&\mathbb{E}_{\tau(\type,\cdot)}\left[U(\cdot,\type)\right]\geq u_\type\\
(\forall\type\in\Types)(\forall\typeb\neq\type)&u_{\typeb}\geq\mathbb{E}_{\tau(\type,\cdot)}\left[U(\belief,\typeb)\right]
\end{array}\right..\nonumber
\end{align}
Then, for each \type, the experiment, $\bsplit(\type,\cdot)$, induces at most $N+M$ posteriors. \label{page-exp}
\end{prop}
The result in \autoref{prop:informed-receiver} affords two simplifications for the designer's problem. First, while the incentive compatibility constraints in \autoref{eq:opt-informed-receiver} impose conditions across the experiments for different types, the optimization problem in \ref{eq:optbis} decouples the problem of designing the experiment for \type\ from the problem of designing the experiment for \typeb. Second, \autoref{prop:informed-receiver} states that each experiment uses at most $N+M$ posteriors. This second simplification is useful when the set of actions available to the agent is rich. Consider, for instance, the case in which $A=\mathbb{R}$ and the agent's payoff is $u(\action,\type,\state)=-(\action-(\state+\type))^2$. In this case, even if the space of actions is a continuum, \autoref{prop:informed-receiver} implies that the designer can focus, without loss of generality, on experiments that induce finitely many beliefs.

\begin{remark}[Connection to the literature]\label{remark:informed-receiver-literature}
In a model where the receiver's payoff depends on the state \state\ only via the posterior mean, \cite{candogan2021optimal} also observe the equivalence between their analogues of program \ref{eq:optbis} and the one in \autoref{eq:opt-informed-receiver}. Because of their assumptions, they model experiments as convex functions and do not rely on the concavification approach. Instead, \autoref{prop:informed-receiver} applies to settings with finitely many states and types, in which the receiver's payoff can depend on the state in arbitrary ways.
\end{remark}
%
{\singlespacing{\bibliographystyle{ecta}
\bibliography{cidt,laura-added}}}
\appendix
\section{Omitted results and proofs of \autoref{sec:main}}\label{appendix:main}

\paragraph{Finite support distributions are without loss in \ref{eq:opt}:} We first show that it is without loss to restrict the choice set in \ref{eq:opt} to distributions with finite support. 
%
%

\begin{prop}[Finite support is without loss of generality]\label{prop:finite-support}
Fix $(\prior,\side)\in\feasible$. Let $\bsplit\in\Delta_{\prior}(\Posteriors)$ be such that it satisfies the constraints. Then, there exists \bsplitopt\ such that $\supp\;\bsplitopt$ is finite and $\mathbb{E}_{\bsplitopt}\left[(\objective,\vectorconstraint)\right]=\mathbb{E}_{\bsplit}\left[(\objective,\vectorconstraint)\right]$. In particular, \bsplitopt\ satisfies the constraints.
\end{prop}
\begin{proof}[Proof of \autoref{prop:finite-support}]
Let \bsplit\ be as in the statement of \autoref{prop:finite-support}. Define:
\begin{align*}
\objectiveopt=\int_{\Posteriors}\objective(\belief)\bsplit(d\belief),
\vectorconstraintopt=\int_{\Posteriors}\vectorconstraint(\belief)\bsplit(d\belief).
\end{align*}
By assumption, $\vectorconstraintopt_\equality=\side_\equality$ and $\vectorconstraintopt_\inequality\geq\side_\inequality$. 

Let $\supportset=\supp\;\bsplit$ denote the support of \bsplit. Let $\convexset=\{(\belief,\objective(\belief),\vectorconstraint(\belief)):\belief\in\supportset
\}$. Because \States\ is finite, the main result in \cite{rubin1958note} implies that $(\prior,\objectiveopt,\vectorconstraintopt)\in\conv\convexset$. Caratheodory's theorem (Theorem 17.1 in \citealp{rockafellar2015convex}) implies that $M\leq N+1+\inequality+\equality$ exists such that $(\prior,\objectiveopt,\vectorconstraintopt)$ can be written as the convex combination of $M$ elements of \convexset. That is, $(\weight_m,\belief_m)_{m=1}^M$ exist such that $(\weight_m)_{m=1}^M\in\Delta_{\prior}\left(\{\belief_1,\dots,\belief_M\}\right)$ and 
\begin{align*}
(\objectiveopt,\vectorconstraintopt)=\sum_{m=1}^M\weight_m(\objective(\belief_m),\vectorconstraint(\belief_m)).
\end{align*}
Letting \bsplitopt\ denote the distribution on \Posteriors\ that assigns probability $\weight_m$ to $\belief_m$, and $0$ otherwise, completes the claim.
\end{proof}

\paragraph{Proof of \autoref{theorem:tomalalemma}:} We first present the statement and proof of Lemma A.1, a step of which later on repeats in the proof of Theorem 3.1:

\begin{lemma}[Relationship between \feasible\ and \constraintset]\label{lemma:convex-domain}
The set \feasible\ is the convex hull of the effective domain of \modifiedobjective. That is, \feasible=\conv\;\constraintset.
\end{lemma}
\begin{proof}[Proof of \autoref{lemma:convex-domain}]
\textbf{\underline{$\feasible\subseteq\conv\left(\constraintset\right)$:}} Let $(\belief,\sideindex)\in\feasible$. By \autoref{prop:finite-support}, $(\weight_m,\belief_m)_{m=1}^M$ exist such that $M\in\naturals$, $\{\belief_1,\dots,\belief_M\}\subseteq\Posteriors$, $(\weight_m)_{m=1}^M\in\Delta_{\belief}(\{\belief_1,\dots,\belief_M\})$, $\sum_{m=1}^M\weight_m\vectorconstraint_\inequality(\belief_m)\geq\sideindex_\inequality$, and $\sum_{m=1}^M\weight_m\vectorconstraint_\equality(\belief_m)=\sideindex_\equality$.

Define $\overline{\side}=\sum_{m=1}^M\weight_m\vectorconstraint(\belief_m)$ and let $\sideindex_m=\vectorconstraint(\belief_m)+\sideindex-\overline{\side}$. Note that
\begin{align}
\sum_{m=1}^M\weight_m\sideindex_m=\sum_{m=1}^M\vectorconstraint(\belief_m)+\sideindex-\overline{\side}=\sideindex.\nonumber
\end{align}
Furthermore, note that $\overline{\side}_\inequality\geq\sideindex_\inequality$, so that $\vectorconstraint_\inequality(\belief_m)\geq\sideindex_{m,\inequality}$. Similarly, $\overline{\side}_\equality=\sideindex_\equality$, so that $\vectorconstraint_\equality(\belief_m)=\sideindex_{m,\equality}$. It thus follows that for all $m\in\{1,\dots,M\}$, $(\belief_m,\sideindex_m)\in\constraintset$ and hence $(\belief,\sideindex)\in\conv\left(\constraintset\right)$.
\paragraph{\underline{$\conv\left(\constraintset\right)\subseteq\feasible$:}} Let $(\belief,\sideindex)\in\conv\left(\constraintset\right)$. Then, $(\weight_m,\belief_m,\sideindex_m)_{m=1}^M$ exist such that $M\in\naturals$, $\{\belief_1,\dots,\belief_M\}\subseteq\Posteriors$, $\{\sideindex_1,\dots,\sideindex_M\}\subset\reals^{\inequality+\equality}$, $(\weight_m)_{m=1}^M\in\Delta_{(\belief,\sideindex)}\left(\{(\belief_1,\sideindex_1),\dots,(\belief_M,\sideindex_M)\}\right)$ and for all $m\in\{1,\dots,M\}$, $\vectorconstraint_\inequality(\belief_m)\geq\sideindex_{m,\inequality}$ and $\vectorconstraint_\equality(\belief_m)=\sideindex_{m,\equality}$. Noting that
\[\sum_{m=1}^M\weight_m\vectorconstraint_\inequality(\belief_m)\geq\sum_{m=1}^M\weight_m\sideindex_{m,\inequality}=\sideindex_\inequality\text{ and }\sum_{m=1}^M\weight_m\vectorconstraint_\equality(\belief_m)=\sum_{m=1}^M\weight_m\sideindex_{m,\equality}=\sideindex_\equality,\]
we obtain that $(\belief,\sideindex)\in\feasible$. 
\end{proof}
The proof of \autoref{theorem:tomalalemma} relies on the following representation of the \cav\ operator:
\begin{prop}[Corollary 17.1.5 in \citealp{rockafellar2015convex}]\label{prop:conv-hull-new}
Given a function $\genericf:\reals^\dimension\mapsto\reals$, we have that for all \variable,
\small
\begin{align*}
\left(\cav\;\genericf\right)(\variable)=\sup\left\{\sum_{m=1}^M\weight_m\genericf(\variable_m)|\{x_1,\dots,x_M\}\subset\dom\;\genericf,(\weight_m)_{m=1}^M\in\Delta_\variable(\{\variable_1,\dots,\variable_M\}),M\leq \dimension+1\right\}.
\end{align*}
\normalsize
\end{prop}
\begin{proof}[Proof of \autoref{theorem:tomalalemma}]
Consider first $(\prior,\side)\notin\feasible$. \autoref{lemma:convex-domain} implies that $\cav\;\modifiedobjective(\prior,\side)=-\infty$, whereas $\Value(\prior,\side)=-\infty$. Thus, from now on, we consider $(\prior,\side)\in\feasible$. 

By \autoref{prop:finite-support} and the definition of $\Value(\prior,\side)$, for each $\epsilon>0$, $(\weight_m,\belief_m)_{m=1}^M$ exist such that the constraints in \ref{eq:opt} are satisfied and $\sum_{m=1}^M\weight_m\objective(\belief_m)\geq\Value(\prior,\gamma)-\epsilon$. The proof of \autoref{lemma:convex-domain} implies that $(\sideindex_m)_{m=1}^M$ exist such that 
\begin{align*}
\sum_{m=1}^M\weight_m\sideindex_m=\sideindex,
\end{align*}
and for all $m\in\{1,\dots,M\}$, $\vectorconstraint_\inequality(\belief_m)\geq\sideindex_{m,\inequality}$ and $\vectorconstraint_\equality(\belief_m)=\sideindex_{m,\equality}$. Applying the definition of \cav\;\modifiedobjective\ in \autoref{prop:conv-hull-new}, we conclude that:
\begin{align*}
\cav\;\modifiedobjective(\prior,\gamma)\geq\sum_{m=1}^M\weight_m\objective(\belief_m)\geq\Value(\prior,\side)-\epsilon.
\end{align*}
Since this holds for all $\epsilon>0$, we obtain that $\cav\;\modifiedobjective(\prior,\gamma)\geq\Value(\prior,\side)$.

By definition of $\cav\;\modifiedobjective(\prior,\gamma)$, for all $\epsilon>0$, $(\weight_m,\belief_m,\sideindex_m)_{m=1}^M$ exist such that (i) $\newline(\weight_m)_{m=1}^M\in\Delta_{(\prior,\side)}\left(\{(\belief_1,\sideindex_1),\dots,(\belief_M,\sideindex_M)\}\right)$, (ii) for all $m\in\{1,\dots, M\}$, $\vectorconstraint_\inequality(\belief_m)\geq\sideindex_{m,\inequality}$, $\vectorconstraint_\equality(\belief_m)=\sideindex_{m,\equality}$, and (iii) $\sum_{m=1}^M\weight_m\modifiedobjective(\belief_m,\sideindex_m)=\sum_{m=1}^M\weight_m\objective(\belief_m)\geq\cav\;\modifiedobjective(\prior,\gamma)-\epsilon$. 

Similar steps to those in the proof of \autoref{lemma:convex-domain} imply that $\Value(\prior,\side)\geq\cav\;\modifiedobjective(\prior,\side)-\epsilon$. Since this holds for all such $\epsilon>0$, we conclude that $\Value(\prior,\side)\geq\cav\;\modifiedobjective(\prior,\side)$.
\end{proof}
\begin{proof}[Proof of \autoref{cor:tomalalemmadrop}]
Under the assumptions of \autoref{cor:tomalalemmadrop}, \Value(\prior,\side) equals
\begin{align}
\Value_\binding(\prior,\side_\binding,\side_\equality)&=\sup_{\bsplit\in\Delta_{\prior}(\Posteriors)}\{\mathbb{E}_{\bsplit}\left[\objective(\belief)\right]|\mathbb{E}_\bsplit\left[\vectorconstraint_\binding(\belief)\right]\geq\side_\binding,\mathbb{E}_\bsplit\left[\vectorconstraint_\equality\right]=\side_\equality\},
\end{align}
 where $(\vectorconstraint_\binding,\side_{\binding})$ is the projection of vector $(\vectorconstraint_\inequality,\side_\inequality)$ on the set $\binding$ of binding constraints. Furthermore, by assumption, a solution exists that attains $\Value_\binding(\cdot)$. \autoref{theorem:tomalalemma} implies that $\Value_\binding(\prior,\side_\binding,\side_\equality)=\cav\objective^{(\vectorconstraint_\binding,\vectorconstraint_\equality)}(\prior,\side_\binding,\side_\equality)$. Thus, a solution to \ref{eq:opt} exists which uses at most $N+\binding+\equality$ beliefs. 
\end{proof}
The proof of \autoref{prop:lagrangian} uses the notion of the concave closure of a function \genericf:
\begin{definition}[Concave closure of \genericf] Given a function $\genericf:\reals^\dimension\mapsto\reals$, the concave closure of \genericf, denoted by $\clcav\;\genericf$, is the function from $\reals^\dimension$ to $\reals$ which satisfies:
\begin{align}\label{eq:cl-cav-hull-h}\tag{$\overline{\cav}\;\genericf$}
\hypograph\left(\clcav\;\genericf\right)=\closure\left(\conv\left(\hypograph\;\genericf\right)\right).
\end{align}
\end{definition}
That is, \clcav\,\genericf\ is the smallest upper-semicontinuous concave function that majorizes \genericf\ (page 36 in \citealp{rockafellar2015convex}). \autoref{example:cav-not-clcav} in \autoref{appendix:examples} illustrates that in general $\clcav\;\genericf$ and $\cav\;\genericf$ do not coincide. However, because the effective domain of \cav\;\genericf\ is a convex set, \cav\;\genericf\ and \clcav\;\genericf\ may only differ on the boundary of $\dom\left(\cav\;\genericf\right)$ (Theorem 7.4 in \citealp{rockafellar2015convex}).
\begin{proof}[Proof of \autoref{prop:lagrangian}]
To prove that \autoref{eq:weak-duality-cav} holds, suppose that \ref{eq:opt} is feasible at (\prior,\side) so that $\bsplit\in\Delta_{\prior}\left(\Posteriors\right)$ exists that satisfies the constraints. For all such \bsplit\ and for any $\dual\in\reals_+^\inequality\times\reals^\equality$, the following holds:
\begin{align*}
&\mathbb{E}_\bsplit\left[\objective(\belief)\right]=\mathbb{E}_\bsplit\left[\objective(\belief)+\sum_{k=1}^{\inequality+\equality}\dual_k\const_k(\belief)\right]-\mathbb{E}_\bsplit\left[\sum_{k=1}^{\inequality+\equality}\dual_k\const_k(\belief)\right]\\
&\leq\mathbb{E}_\bsplit\left[\objective(\belief)+\sum_{k=1}^{\inequality+\equality}\dual_k\const_k(\belief)\right]-\sum_{k=1}^{\inequality+\equality}\dual_k\side_k
\leq\cav\left(\objective+\sum_{k=1}^{\inequality+\equality}\dual_k\const_k\right)(\prior)-\sum_{k=1}^{\inequality+\equality}\dual_k\side_k,
\end{align*}
where the first inequality follows from $\bsplit$ being a feasible solution for \ref{eq:opt} and $\dual_\inequality\geq0$ and the second inequality follows because $\bsplit\in\Delta_{\prior}(\Posteriors)$. We conclude that for any $\dual\in\reals_+^\inequality\times\reals^\equality$, 
\begin{align}\label{eq:almost-weak-duality}
\Value(\prior,\side)=\sup_{\bsplit\in\Delta_{\prior}(\Posteriors):\mathbb{E}_\bsplit\vectorconstraint_\inequality\geq\side_\inequality,\mathbb{E}_\bsplit\vectorconstraint_\equality\geq\side_\equality}\mathbb{E}_\bsplit\left[\objective\right]\leq\cav\left(\objective+\sum_{k=1}^{\inequality+\equality}\dual_k\const_k\right)(\prior)-\sum_{k=1}^{\inequality+\equality}\dual_k\side_k.
\end{align}
Since \autoref{eq:almost-weak-duality} holds for every $\dual\in\reals_+^\inequality\times\reals^\equality$, taking the infimum on both sides delivers \autoref{eq:weak-duality-cav}.

We now prove that \autoref{eq:no-duality-gap} holds for (\prior,\side) in the interior of \feasible. Let $\langle\variable,\level\rangle$ denote the inner product, that is for vectors $\variable,\level\in\reals^\dimension$, $\langle\variable,\level\rangle=\sum_{i=1}^{\dimension}\variable_i\level_i$. For a function $\genericf:\reals^\dimension\mapsto\reals$, let $\genericf\conjugate$ denote its Fenchel conjugate, where $\genericf\conjugate(\dual)=\sup_\variable\{\langle\variable,\dual\rangle-\genericf(\variable)\}$. Theorem 11.1 in \cite{rockafellar2009variational} implies $-\clcav(-\genericf)(\variable)=(\genericf\conjugate)\conjugate(\variable)$ whenever $-\cav(-\genericf)$ is proper. Hence,
\begin{align}\label{eq:concav-fenchel}
\clcav\;\genericf(\variable)=\inf_\price\left\{\langle\variable,\price\rangle+\sup_{\variableindex}(\genericf(\variableindex)-\langle\price,\variableindex\rangle)\right\}.
\end{align}
We now apply this to the function $\modifiedobjective$ at $(\prior,\side)$ in the interior of \feasible; in particular, note that this implies that \constraintset\ is non-empty. Let $\price_{\prior}\in\reals^N$ and $\price_\side\in\reals^{\inequality+\equality}$, 
\begin{align}\label{eq:clcav-modifiedg}
&\clcav\;\modifiedobjective(\prior,\side)=\inf_{(\price_{\prior},\price_\side)\in\reals^{N+\inequality+\equality}}\left\{\langle\price_{\prior},\prior\rangle+\langle\price_\side,\side\rangle+\sup_{\belief,\sideindex}\left(\modifiedobjective(\belief,\sideindex)-\langle\price_{\prior},\belief\rangle-\langle\price_\side,\sideindex\rangle\right)\right\}\nonumber
\\
&=\inf_{(\price_{\prior},\price_\side)\in\reals^{N+\inequality+\equality}}\left\{\langle\price_{\prior},\prior\rangle+\langle\price_\side,\side\rangle+\sup_{(\belief,\sideindex)\in C}\left(\objective(\belief)-\langle\price_{\prior},\belief\rangle-\langle\price_\side,\sideindex\rangle\right)\right\}.
\end{align}
Because \constraintset\ is non-empty, the supremum in the second line is not $-\infty$. If  $k\in\{1,\dots,\inequality\}$ exists such that $\price_{\side,k}>0$, then letting $\sideindex_k\rightarrow-\infty$, the supremum is $+\infty$. We can thus restrict attention to $\price_{\side,k}\leq0$ for $k\in\{1,\dots,\inequality\}$. Setting $\dual_{\side}=-\price_\side$ we get,
\begin{align}\label{eq:clcav-lagrangefg}
&\clcav\;\modifiedobjective(\prior,\side)=\inf_{\price_{\prior}\in\reals^N,\dual_\side\in\reals_+^\inequality\times\reals^\equality}\left\{\langle\price_{\prior},\prior\rangle-\langle\dual_\side,\side\rangle+\sup_{(\belief,\sideindex)\in C}\left(\objective(\belief)-\langle\price_{\prior},\belief\rangle+\langle\dual_\side,\sideindex\rangle\right)\right\}\nonumber
\\
&=\inf_{\price_{\prior}\in\reals^N,\dual_\side\in\reals_+^\inequality\times\reals^\equality}\left\{\langle\price_{\prior},\prior\rangle-\langle\dual_\side,\side\rangle+\sup_{\belief}\left(\objective(\belief)-\langle\price_{\prior},\belief\rangle+\sum_{k=1}^{\inequality+\equality}\dual_{\side,k}\const_k(\belief)\right)\right\}\\
&=\inf_{\dual_\side\in\mathbb{R}_+^\inequality\times\mathbb{R}^{\equality}}\left\{\inf_{\price_{\prior}\in\reals^N}\left[\langle\price_{\prior},\prior\rangle+\sup_{\belief}\left(\objective(\belief)+\sum_{k=1}^{\inequality+\equality}\dual_{\side,k}\const_k(\belief)-\langle\price_{\prior},\belief\rangle\right)\right]-\langle\dual_\side,\side\rangle\right\}\nonumber\\
&=\inf_{t_\side\in\mathbb{R}_+^\inequality\times\mathbb{R}^{\equality}}\left\{\clcav(\objective+\sum_{k=1}^{\inequality+\equality}\dual_{\side,k}\const_k)(\prior)-\langle\dual_\side,\side\rangle\right\}=\inf_{t_\side\in\mathbb{R}_+^\inequality\times\mathbb{R}^{\equality}}\left\{\cav(\objective+\sum_{k=1}^{\inequality+\equality}\dual_{\side,k}\const_k)(\prior)-\langle\dual_\side,\side\rangle\right\},\nonumber
\end{align}
where (i) the second equality follows from noticing that $\sideindex_k=\const_k(\belief)$, whenever $k>\inequality$, while it is optimal to set $\sideindex_k=\const_k(\belief)$ whenever $k\leq \inequality$ since $\dual_{\side,k}\geq 0$,
(ii) the third equality is just a rewriting,  (iii) the fourth equality follows from noting that the infimum in the square brackets is the definition of the concave closure of $\objective+\sum \dual_{\side,k}\const_k$ (see \autoref{eq:concav-fenchel}), and (iv) the final equality follows from noting that (\prior,\side) in the interior of \feasible\ implies that \prior\ is in the interior of \Posteriors, and the concave closure coincides with the concave hull in the interior of effective domain of $\objective+\sum_{k=1}^{\inequality+\equality}\dual_k\const_k$, which is \Posteriors\ by \autoref{assumption:main}.

\autoref{eq:no-duality-gap} then follows by noting that \cav\;\modifiedobjective\ coincides with \clcav\;\modifiedobjective\ in the interior of \feasible\  (Theorem 7.4 in \citealp{rockafellar2015convex}).

Finally, to see that the third statement of \autoref{prop:lagrangian} holds, note the following. Similar steps to those in the proof of Observations \ref{observation:f-closed} and \ref{observation:solution} show that under Assumptions \hyperlink{assumptionf}{\objective} and \hyperlink{assumptiong}{\vectorconstraint}, the correspondence that assigns to each $(\prior,\side)\in\feasible$ the subset of $\Delta_{\prior}(\Posteriors)$ that satisfies the constraints is a non-empty, compact-valued, upper-hemicontinuous correspondence. A version of the Theorem of the Maximum (Lemma 17.30 in \citealp{aliprantisborder}) implies that \Value(\prior,\side) is upper-semicontinuous. \autoref{theorem:tomalalemma} then implies that \cav\;\modifiedobjective\ is upper-semicontinuous and hence, coincides with \clcav\;\modifiedobjective. 
%
\end{proof}
\begin{proof}[Proof of \autoref{theorem:multiplier-existence}]\label{appendix:multiplier-existence}

Let $\jointcav_\feasibleset=\{\level\in\jointcav|\level_\inequality\geq \side_\inequality,\level_\equality=\side_\equality\}$ denote the elements in \jointcav\ that satisfy the constraints in \ref{eq:opt} at (\prior,\side). In the supplementary material \citep{supplement}, we show that $\jointcav_\feasibleset\neq\emptyset$ if and only if $(\prior,\side)\in\feasible$, and that $\sup\{\level_\objective|\level\in\jointcav_\feasibleset\}=\Value(\prior,\side)$.

Let $\canonicalobj\in\reals^{\inequality+\equality+1}$ denote the canonical vector that has a 1 in the first coordinate and 0 otherwise, whereas for $i\in\{1,\dots,\inequality+\equality\}$ $\canonicalconst$ is the canonical vector that has a 1 in the $(i+1)^{th}$ coordinate and $0$ otherwise. Finally, let $k_0$ such that $k_0(\level)=\langle\canonicalobj,\level\rangle=\level_\objective$ if $\level\in\jointcav$ and $-\infty$ otherwise.

The problem $\sup\{\level_\objective|\level\in\jointcav_\feasibleset\}$, whose value coincides with that of \ref{eq:opt}, corresponds to the following finite-dimensional optimization problem on the convex set \jointcav:
\begin{align}\tag{\OPT$_\jointcav$}\label{eq:opt-jointcav}
&\sup_{\level\in\jointcav} k_0(\level)\\
\text{ s.t. }&\left\{\begin{array}{l}(\forall i\in\{1,\dots,\inequality\})\langle\canonicalconst,\level\rangle\geq\side_\inequality \\ (\forall i\in\{\inequality+1,\dots,\inequality+\equality\})\langle\canonicalconst,\level\rangle=\side_\equality\end{array}\right..\nonumber
\end{align}
Assuming \ref{eq:opt} satisfies \hyperlink{assumptioncs}{Assumption S} at (\prior,\side), Corollary 28.2.2 in \cite{rockafellar2015convex} implies that  $\dual\conjugate\in\reals_+^\inequality\times\reals^\equality$ exists such that  the supremum of the function
\[\lagrangian(\level,\dual\conjugate)=k_0(\level)+\sum_{i=1}^{\inequality+\equality}\dual_i\conjugate\left(\langle\canonicalconst,\level\rangle-\side_i\right),\]
whose effective domain is \jointcav, is finite and equal to the value of \ref{eq:opt-jointcav}. That is, \dual\conjugate\ is a Lagrange multiplier for \ref{eq:opt-jointcav}.
Since the value of \ref{eq:opt-jointcav} is the value of \ref{eq:opt}, we have the following:
\begin{align}
&\Value(\prior,\side)=\sup_{y\in\jointcav}\left[k_0(\level)+\sum_{i=1}^{\inequality+\equality}\dual_i\conjugate\left(\langle\canonicalconst,\level\rangle-\side_i\right)\right]=\sup_{\level\in\jointcav}\left[\level_\objective+\sum_{i=1}^{\inequality+\equality}\dual_i\conjugate\left(\level_i-\side_i\right)\right]\nonumber\\
&=\sup_{\tau\in\Delta_{\prior}(\Posteriors)}\left[\mathbb{E}_\bsplit\left[\objective\right]+\sum_{i=1}^{\inequality+\equality}\dual_i\conjugate\left(\mathbb{E}_\bsplit[\const_i]-\side_i\right)\right]=\sup_{\tau\in\Delta_{\prior}(\Posteriors)}\mathbb{E}_\bsplit\left[\objective+\sum_{i=1}^{\inequality+\equality}\dual_i\conjugate\const_i\right]-\sum_{i=1}^{\inequality+\equality}\dual_i\conjugate\side_i\nonumber\\
&=\cav\left[\objective+\sum_{i=1}^{\inequality+\equality}\dual_i\conjugate\const_i\right](\prior)-\sum_{i=1}^{\inequality+\equality}\dual_i\conjugate\side_i,\nonumber
\end{align}
where the first equality is by definition of \dual\conjugate, the second is obtained by spelling out the inner products, the third is by definition of $\level\in\jointcav$, the fourth groups all terms that correspond to \bsplit, and the last is by definition of the concave hull at the prior.
\end{proof}

\begin{proof}[Proof of \autoref{prop:agreement}]
Because \objective\ is upper-semicontinuous, a solution \bsplitopt\ exists that attains \cav\;\objective. Furthermore, Corollary 17.1.5 in \cite{rockafellar2015convex} implies that without loss of generality the cardinality of the support of \bsplitopt\ is at most $N$.

Let \bsplit\ be such that $\bsplit\in\Delta_{\prior}(\Posteriors)$ and $\bsplit$ satisfies the constraints of \ref{eq:opt}. Note that \objective\ (weakly) prefers \bsplitopt\ to \bsplit. Since $(\objective,\vectorconstraint)$ are in agreement, we have the following:
\begin{align*}
\mathbb{E}_{\bsplitopt}\left[\vectorconstraint_\inequality\right]\geq\mathbb{E}_{\bsplit}\left[\vectorconstraint_\inequality\right]&\geq\side_\inequality,\;\;
\mathbb{E}_{\bsplitopt}\left[\vectorconstraint_\equality\right]=\mathbb{E}_{\bsplit}\left[\vectorconstraint_\equality\right]=\side_\equality.
\end{align*}
Thus, \bsplitopt\ is feasible for \ref{eq:opt}. \autoref{prop:agreement} then follows.
\end{proof}
\begin{proof}[Proof of \autoref{prop:gid}]
Let $\bsplitopt$ denote a solution to \ref{eq:opt} and suppose that $|\support\bsplitopt|=M>K$. Then, there exists $k\in\{1,\dots,K\}$ such that $|\Delta_k\cap\support\bsplitopt|\geq 2$. Letting $\support\bsplitopt=\{\beliefopt_1,\dots,\beliefopt_M\}$, define $\bsplitb_k=\sum_{m=1}^M\bsplitopt(\beliefopt_m)\mathbbm{1}[\beliefopt_m\in\Delta_k]$ and let $\beliefoptb_k=\sum_{m=1}^M\nicefrac{\bsplitopt(\beliefopt_m)}{\bsplitb_k}\beliefopt_m\mathbbm{1}[\beliefopt_m\in\Delta_k]$.

Consider the distribution over posteriors \bsplitb\ that places weight $\bsplitopt(\beliefopt_m)$ on $\beliefopt_m$ whenever $\beliefopt_m\in\{\beliefopt_1,\dots,\beliefopt_M\}\setminus\Delta_k$, places weight $\bsplitb_k$ on $\beliefoptb_k$, and $0$ otherwise. It is immediate to check that it averages out to the prior. 

Since \objective\ and $\vectorconstraint_\inequality$ are concave over $\Delta_k$, then $\mathbb{E}_{\bsplitb}[(\objective,\vectorconstraint_\inequality)]\geq\mathbb{E}_{\bsplitopt}[(\objective,\vectorconstraint_\inequality)]\geq\left(\mathbb{E}_{\bsplitopt}[\objective],\side_\inequality\right)$. Similarly, because $\vectorconstraint_\equality$ is affine on $\Delta_k$, we have that $\mathbb{E}_{\bsplitb}[\vectorconstraint_\equality]=\mathbb{E}_{\bsplitopt}[\vectorconstraint_\equality]=\side_\equality$. Thus, \bsplitb\ is feasible and (weakly) increases the objective. \autoref{prop:gid} follows.
\end{proof}
\section{Proofs of \autoref{sec:limited-commitment}}\label{appendix:limited-commitment}
\begin{proof}[Proof of \autoref{prop:med}]
Consider the following program: 
\small
\begin{align}\label{eq:programadj}\tag{$\mathcal{A}$}
&\max_{\beta:\Types\mapsto\Delta\Posteriors,\alpha:\Posteriors\mapsto\Delta(\Allocations_1)}\sum_{\state\in\States}\prior(\state)\mathbb{E}_{\beta(\cdot|\type)}\left[\mathbb{E}_{\alpha(\cdot|\posterior)}\left[v(\allocation_1,\allocation_2,\state)\right]\right]\\
&\text{s.t.}\left\{\begin{array}{ll}&\mathbb{E}_{\beta(\cdot|\state_1)}\left[\mathbb{E}_{\alpha(\cdot|\posterior)}\left[u(\allocation_1,\allocation_2(\allocation_1,\posterior),\state_1)\right]\right]\geq0\\
(\forall i\in\{2,\dots,N\})&\mathbb{E}_{\beta(\cdot|\state_i)-\beta(\cdot|\state_{i-1})}\left[\mathbb{E}_{\alpha(\cdot|\posterior)}\left[u(\allocation_1,\allocation_2(\allocation_1,\posterior),\state_i)\right]\right]\geq0\\
(\forall i\in\{1,\dots,N-1\})&\mathbb{E}_{\beta(\cdot|\state_i)-\beta(\cdot|\state_{i+1})}\left[\mathbb{E}_{\alpha(\cdot|\posterior)}\left[u(\allocation_1,\allocation_2(\allocation_1,\posterior),\state_i)\right]\right]\geq0
\end{array}\right..\nonumber
\end{align}
\normalsize
We show that the solution to \ref{eq:programadj} satisfies all the constraints of \ref{eq:opt-limited-commitment}. To simplify notation, in what follows, let
\begin{align*}
u(\belief,\state_i)=\mathbb{E}_{\alpha(\cdot|\posterior)}\left[u(\allocation_1,\allocation_2(\allocation_1,\belief),\state_i)\right].
\end{align*}
Note first that the solution to \ref{eq:programadj} satisfies that for all $i\geq 2$,
\begin{align*}
\mathbb{E}_{\beta(\cdot|\state_i)}\left[u(\posterior,\state_i)\right]&\geq\mathbb{E}_{\beta(\cdot|\state_{i-1})}\left[u(\posterior,\state_i)\right]\\
\mathbb{E}_{\beta(\cdot|\state_{i-1})}\left[u(\posterior,\state_{i-1})\right]&\geq\mathbb{E}_{\beta(\cdot|\state_{i})}\left[u(\posterior,\state_{i-1})\right],
\end{align*}
%
so that for all $i\geq 2$, we have
\begin{align}\label{eq:direction}
\mathbb{E}_{\beta(\cdot|\state_i)-\beta(\cdot|\state_{i-1})}\left[u(\belief,\state_i)-u(\belief,\state_{i-1})\right]\geq 0.
\end{align}
\autoref{eq:direction} has two implications. First, \autoref{definition:med} together with equation \autoref{eq:direction} implies that if $k<i$, then it cannot be the case that
\begin{align*}
\mathbb{E}_{\beta(\cdot|\state_k)-\beta(\cdot|\state_{k-1})}\left[u(\belief,\state_i)-u(\belief,\state_{k})\right]<0.
\end{align*}
Hence, we must have $\mathbb{E}_{\beta(\cdot|\state_k)-\beta(\cdot|\state_{k-1})}u(\belief,\state_i)\geq\mathbb{E}_{\beta(\cdot|\state_k)-\beta(\cdot|\state_{k-1})}u(\belief,\state_k)$, when $k<i$. Second, if $k>i$, \autoref{eq:direction} evaluated at $k$ together with monotonic expectational differences implies that $\mathbb{E}_{\beta(\cdot|\state_k)-\beta(\cdot|\state_{k-1})}u(\belief,\state_k)\geq\mathbb{E}_{\beta(\cdot|\state_k)-\beta(\cdot|\state_{k-1})}u(\belief,\state_i)$. We use these two implications in what follows.

To show that the statement of the proposition holds, consider $i$ and $j<i-1$. The solution to \ref{eq:programadj} satisfies that for $k\in\{j+1,\dots,i\}$:
\begin{align*}
\mathbb{E}_{\beta(\cdot|\state_k)}[u(\belief,\state_k)]&\geq\mathbb{E}_{\beta(\cdot|\state_{k-1})}[u(\belief,\state_k)].
\end{align*}
Adding up over $k\in\{j+1,\dots,i\}$, we obtain
\begin{align}\label{eq:dicimpl}
\sum_{k=j+1}^i\mathbb{E}_{(\beta(\cdot|\state_k)-\beta(\cdot|\state_{k-1}))}u(\belief,\state_k)\geq0.
\end{align}
As discussed above, \autoref{definition:med} together with \autoref{eq:direction} imply the left-hand side of \autoref{eq:dicimpl} is bounded above by
\begin{align}\label{eq:medimpl}
\sum_{k=j+1}^i\mathbb{E}_{(\beta(\cdot|\state_k)-\beta(\cdot|\state_{k-1}))}\left[u(\belief,\state_i)\right]=\mathbb{E}_{\beta(\cdot|\state_i)-\beta(\cdot|\state_j)}\left[u(\belief,\state_i)\right].
\end{align}
Equations \ref{eq:dicimpl} and \ref{eq:medimpl} imply
\begin{align*}
\mathbb{E}_{\beta(\cdot|\state_i)}[u(\belief,\state_i)]\geq\mathbb{E}_{\beta(\cdot|\state_j)}[u(\belief,\state_i)].
\end{align*}
Therefore, the constraint that $i$ does not report $j<i-1$ holds. Similarly, consider $i$ and $j>i+1$. The solution to \ref{eq:programadj} satisfies that for $k\in\{i,\dots,j-1\}$
\begin{align*}
\mathbb{E}_{\beta(\cdot|\state_k)}[u(\belief,\state_k)]&\geq\mathbb{E}_{\beta(\cdot|\state_{k+1})}[u(\belief,\state_k)].
\end{align*}
Adding up over $k\in\{i,\dots,j-1\}$, we obtain
\begin{align}\label{eq:dicimpl2}
\sum_{k=i}^{j-1}\mathbb{E}_{(\beta(\cdot|\state_k)-\beta(\cdot|\state_{k+1}))}u(\belief,\state_k)\geq0.
\end{align}
As discussed above, \autoref{definition:med} together with \autoref{eq:direction} imply that the left-hand side is bounded above by
\begin{align}\label{eq:medimpl2}
\sum_{k=i}^{j-1}\mathbb{E}_{(\beta(\cdot|\state_k)-\beta(\cdot|\state_{k+1}))}u(\belief,\state_i)=\mathbb{E}_{(\beta(\cdot|\state_i)-\beta(\cdot|\state_{j}))}u(\belief,\state_i).
\end{align}
\autoref{eq:medimpl2} follows because \autoref{eq:direction} implies $\mathbb{E}_{(\beta(\cdot|\state_k)-\beta(\cdot|\state_{k+1}))}u(\belief,\state_k)$ is decreasing in $k$. 

Equations \ref{eq:dicimpl2} and \ref{eq:medimpl2} imply
\begin{align*}
\mathbb{E}_{\beta(\cdot|\state_i)}[u(\belief,\state_i)]\geq\mathbb{E}_{\beta(\cdot|\state_j)}[u(\belief,\state_i)].
\end{align*}
Therefore, the incentive constraint that $i$ does not report $j$, $j>i+1$ holds. 

Finally, because we have all incentive compatibility constraints, it follows that, when $u_i$ satisfies \autoref{definition:med}, the participation constraints for $i\geq 2$ are implied by the participation constraint for $i=1$. To see this, note the following. First, because all incentive compatibility constraints are satisfied, we have that for all $i\geq 2$,
\begin{align}\label{eq:inot1}
\mathbb{E}_{\beta(\cdot|\state_i)}[u(\belief,\state_i)]\geq\mathbb{E}_{\beta(\cdot|\state_1)}[u(\belief,\state_i)].
\end{align}
\textcolor{black}{We can write the right-hand side of \autoref{eq:inot1} as}
\begin{align}
\mathbb{E}_{\beta(\cdot|\state_1)}[u(\belief,\state_1)+u(\belief,\state_i)-u(\belief,\state_1)].
\end{align}
\textcolor{black}{Now, since the family $\{u(\cdot,\state):\state\in\States\}$ satisfies \autoref{definition:med}, we have that}
\begin{align*}
u(\belief,\state_i)-u(\belief,\state_1)=(b(\state_i)-b(\state_1))\mathbb{E}_{\alpha(\cdot|\posterior)}[h_1(\allocation_1,\allocation_2(\allocation_1,\belief))]+c(\state_i)-c(\state_1).
\end{align*}
Moreover, recall from footnote \ref{ftn:participation}, that we assume that $u(\allocation_1^*,\allocation_2^*,\state_i)=0$ for all $i\in\{1,\dots,N\}$. Hence, we can rewrite the above as:
\begin{align*}
&u(\belief,\state_i)-u(\belief,\state_1)=(b(\state_i)-b(\state_1))\mathbb{E}_{\alpha(\cdot|\posterior)}[h_1(\allocation_1,\allocation_2(\allocation_1,\belief))]+c(\state_i)-c(\state_1)
\\&=(b(\state_i)-b(\state_1))(\mathbb{E}_{\alpha(\cdot|\posterior)}[h_1(\allocation_1,\allocation_2(\allocation_1,\belief))]-h_1(\allocation_1^*,\allocation_2^*))\geq 0,
\end{align*}
since $b(\cdot)$ is increasing in $\state_i$ and $h_1$ is minimized at $(\allocation_1^*,\allocation_2^*)$ by the assumption in footnote \ref{ftn:participation}.
\end{proof}
\begin{proof}[Proof of \autoref{cor:bound-limited-commitment-1}]
\autoref{prop:med} implies that under monotonic expectational differences, it is enough to consider the solution to \ref{eq:programadj}. Writing it in terms of the distribution of posteriors it induces, we obtain:
\begin{align}\tag{$\mathcal{A}^\prime$}
&\max_{\alpha:\Posteriors\mapsto\Delta(\Allocations_1),\allocation_2\in\bestresponse_2}\max_{\tau\in\Delta_{\prior}\Posteriors}\mathbb{E}_\tau\left[\mathbb{E}_{\alpha(\cdot|\belief)}\left[\sum_{\state\in\States}\belief(\state)v(\allocation_1,\allocation_2(\allocation_1,\belief),\state)\right]\right]\\
&s.t.\left\{
\begin{array}{l}
\mathbb{E}_{\tau(\cdot)}\left[\mathbb{E}_{\alpha(\cdot|\belief)}\left[\frac{\belief(\state)}{\prior(\state)}u(\allocation_1,\allocation_2(\allocation_1,\belief),\state_1)\right]\right]\geq 0\\
(\forall i\in\{2,\dots,N\})\mathbb{E}_{\tau(\cdot)}\left[\mathbb{E}_{\alpha(\cdot|\belief)}\left[\left(\frac{\belief(\state_i)}{\prior(\state_i)}-\frac{\belief(\state_{i-1})}{\prior(\state_{i-1})}\right)u(\allocation_1,\allocation_2(\allocation_1,\belief),\state_i)\right]\right]\geq0\\
(\forall i\in\{1,\dots,N-1\})\mathbb{E}_{\tau(\cdot)}\left[\mathbb{E}_{\alpha(\cdot|\belief)}\left[\left(\frac{\belief(\state_i)}{\prior(\state_i)}-\frac{\belief(\state_{i+1})}{\prior(\state_{i+1})}\right)u(\allocation_1,\allocation_2(\allocation_1,\belief),\state_i)\right]\right]\geq0\end{array}\right..\nonumber
\end{align}

Fix $\alpha:\Posteriors\mapsto\Delta(\Allocations_1)$ and a selection $\allocation_2(\cdot)\in\bestresponse_2$, and consider the program:
\begin{align}\tag{$\mathcal{A}_\alpha^\prime$}\label{eq:programadjall}
&\max_{\tau\in\Delta_{\prior}\left(\Posteriors\right)}\mathbb{E}_\tau\left[\mathbb{E}_{\alpha(\cdot|\belief)}\left[\sum_{\state\in\States}\belief(\state)v(\allocation_1,\allocation_2(\allocation_1,\belief),\state)\right]\right]\\
&s.t.\left\{
\begin{array}{l}
\mathbb{E}_{\tau(\cdot)}\left[\mathbb{E}_{\alpha(\cdot|\belief)}\left[\frac{\belief(\state_1)}{\prior(\state_1)}u(\allocation_1,\allocation_2(\allocation_1,\belief),\state_1)\right]\right]\geq 0\\
(\forall i\in\{2,\dots,N\})\mathbb{E}_{\tau(\cdot)}\left[\mathbb{E}_{\alpha(\cdot|\belief)}\left[\left(\frac{\belief(\state_i)}{\prior(\state_i)}-\frac{\belief(\state_{i-1})}{\prior(\state_{i-1})}\right)u(\allocation_1,\allocation_2(\allocation_1,\belief),\state_i)\right]\right]\geq0\\
(\forall i\in\{1,\dots,N-1\})\mathbb{E}_{\tau(\cdot)}\left[\mathbb{E}_{\alpha(\cdot|\belief)}\left[\left(\frac{\belief(\state_i)}{\prior(\state_i)}-\frac{\belief(\state_{i+1})}{\prior(\state_{i+1})}\right)u(\allocation_1,\allocation_2(\allocation_1,\belief),\state_i)\right]\right]\geq0\end{array}\right..\nonumber
\end{align}
Note that there might be allocations $\alpha$ for which there is no $\tau$ that satisfies the incentive compatibility and/or participation constraints. To address this issue, let $C_\alpha$ denote the policies $\tau$ that satisfy the constraints in \ref{eq:programadjall}. Let $f_\alpha^0(\tau)$ denote
\begin{align*}
\mathbb{E}_\tau\left[\mathbb{E}_{\alpha(\cdot|\belief)}\left[\sum_{\state\in\States}\belief(\state)v(\allocation_1,\allocation_2(\allocation_1,\belief),\state)\right]\right],
\end{align*}
and let 
\begin{align*}
f_\alpha(\tau)=\left\{\begin{array}{ll}f_\alpha^0(\tau)&\text{if }\tau\in C_\alpha\\-\infty&\text{otherwise }\end{array}\right..
\end{align*}
In what follows, $f_\alpha(\tau)$ is the objective function under consideration. Note that letting, 
\begin{align*}
g_i(\belief)&=\left(\frac{\posterior(\state_i)}{\prior(\state_i)}-\frac{\posterior(\state_{i+1})}{\prior(\state_{i+1})}\right)u(\posterior,\state_i),\quad i\in\{1,\dots,N-1\}\\
g_{N-2+i}(\belief)&=\left(\frac{\posterior(\state_i)}{\prior(\state_i)}-\frac{\posterior(\state_{i-1})}{\prior(\state_{i-1})}\right)u(\posterior,\state_i),\quad i\in\{2,\dots,N\}\\
g_{2N-1}(\belief)&=\frac{\posterior(\state_1)}{\prior(\state_1)}u(\posterior,\state_1),
\end{align*}
we can write \ref{eq:programadjall} as a special case of \ref{eq:opt}, with $\inequality=2N-1$. \autoref{cor:bound} implies that any finite solution to \ref{eq:programadjall} uses at most $3N-1$ beliefs.
\end{proof}
\begin{proof}[Proof of \autoref{cor:ir-binding}]
Towards a contradiction, suppose the participation constraint of $\state_1$ is not binding. Then, let $\epsilon=\mathbb{E}_{\beta(\cdot|\state_1)}\left[\mathbb{E}_{\alpha(\cdot|\posterior)}[\tilde{u}(\allocationb_1,\allocation_2(\cdot),\state_1)]\right]$. Consider a mechanism that increases all transfers, $\transfer(\posterior)+\epsilon$. All incentive constraints continue to be satisfied, the participation constraint for $\state_1$ binds, and revenue increases, contradicting that the solution was optimal.
\end{proof}

\begin{proof}[Proof of \autoref{prop:vs}]
Given a selection $\allocation_2(\allocationb_1,\posterior)$ from the principal's best response correspondence in period $2$ when his beliefs are \posterior, let 
\begin{align*}
\overline{u}(\allocationb_1,\allocation_2(\allocationb_1,\posterior),\state_i)&=\tilde{u}(\allocationb_1,\allocation_2(\allocationb_1,\posterior),\state_i)\\&-\frac{1-\sum_{n\leq i}\prior(\state_n)}{\prior(\state_i)}(\tilde{u}(\allocationb_1,\allocation_2(\allocationb_1,\posterior),\state_i)-\tilde{u}(\allocationb_1,\allocation_2(\allocationb_1,\posterior),\state_{i-1})).
\end{align*}
Then, replacing the constraints in \ref{eq:relaxed} in the principal's objective function, we obtain the following expression:
\begin{align*}
\mathbb{E}_{\tau}[\mathbb{E}_{\alpha(\cdot|\posterior)}\sum_{\state\in\States}\belief(\state)\left(\tilde{v}(\allocationb_1,\allocation_2(\allocationb_1,\posterior),\state)+\overline{u}(\allocationb_1,\allocation_2(\allocationb_1,\posterior),\state)\right)]\equiv\mathbb{E}_\bsplit\left[\check{v}(\alpha,\allocation_2,\posterior)\right].
\end{align*}
Therefore, we can write \ref{eq:relaxed} as 
\begin{align}\label{eq:relaxed-rewrite}
\max_{\tau\in\Delta_{\prior}(\Posteriors),\alpha,\allocation_2(\cdot)\in\bestresponse_2}\mathbb{E}_\tau\left[\check{v}(\alpha,\allocation_2,\posterior)\right].
\end{align}
That is, the solution to the relaxed problem is obtained by maximizing a version of the virtual surplus, represented by $\check{v}$, and then choosing a distribution over posteriors that averages out to the prior. The following remark is in order:
\begin{remark}[Tie-breaking]\label{remark:tie-breaking}
So far we have remained silent about how $\allocation_2(\allocationb_1,\posterior)$ is chosen, beyond the restriction that $\allocation_2(\cdot)\in\bestresponse_2$. We can use the function $\tilde{v}(\allocationb_1,\allocation_2(\allocationb_1,\posterior),\state_i)+\overline{u}(\allocationb_1,\allocation_2(\allocationb_1,\posterior),\state_i)$ to determine how to break the possible ties in $\bestresponse_2$ and make the principal's objective function upper-semicontinuous. In fact, if $\allocation_2,\allocationb_2\in\bestresponse_2(\allocationb_1,\posterior)$, then in the relaxed program, $\allocation_2$ is selected as long as
\begin{align*}
\sum_{\state\in\States}\posterior(\state)\left[\tilde{v}(\allocationb_1,\allocation_2,\state)+\overline{u}(\allocationb_1,\allocation_2,\state)\right]\geq\sum_{\state\in\States}\posterior(\state)\left[\tilde{v}(\allocationb_1,\allocationb_2,\state)+\overline{u}(\allocationb_1,\allocationb_2,\state)\right].
\end{align*}
In other words, ties are broken in favor of the virtual surplus.
\end{remark}
We now illustrate how to solve the program in \autoref{eq:relaxed-rewrite}. Towards this, fix the selection $\allocation_2^*$ as in \autoref{remark:tie-breaking}. Because the program is separable in the allocation $\alpha$ across posteriors $\posterior$, the solution can be obtained in two steps. First, for each posterior \posterior, we maximize $\check{v}(\cdot,\allocation_2^*,\posterior)$ with respect to $\alpha$. Denote the value of this problem $\hat{v}(\posterior)$. Second, we choose $\tau$ to maximize the expectation of $\hat{v}(\cdot)$ subject to the constraint that $\tau$ is Bayes' plausible. A straightforward application of Corollary 17.1.5 in \cite{rockafellar2015convex} implies that the solution to \ref{eq:relaxed} involves at most $N$ posteriors.
\end{proof}

To introduce \autoref{prop:relaxed-more-conditions}, we construct two objects from the solution to the relaxed program, $\joint\conjugate\equiv(\bsplit\conjugate,\tilde{\alpha}\conjugate,\allocation_2\conjugate)$. First, similar to the commitment solution, we can always find a transfer $\transfer_{\joint\conjugate}(\state_i)$  that solves $\state_i$'s binding constraint for each $i\in\{1,\dots,N\}$. Namely,   using the participation constraint of $\state_1$, let
\begin{align*}
\transfer_{\joint\conjugate}(\state_1)=\mathbb{E}_{\bsplit\conjugate}\left[\mathbb{E}_{\tilde{\alpha}\conjugate}\left[\frac{\belief(\state_1)}{\prior(\state_1)}u(\allocationb_1,\allocation_2\conjugate(\allocationb_1,\belief),\state_1)\right]\right].
\end{align*}
Recursively, using the downward looking incentive constraint of $\state_i$ and the transfer $\transfer_{\joint\conjugate}(\state_{i-1})$, define:
\begin{align*}
\transfer_{\joint\conjugate}(\state_i)=\mathbb{E}_{\bsplit\conjugate}\left[\mathbb{E}_{\tilde{\alpha}\conjugate}\left[\left(\frac{\belief(\state_i)}{\prior(\state_i)}-\frac{\belief(\state_{i-1})}{\prior(\state_{i-1})}\right)u(\allocationb_1,\allocation_2\conjugate(\allocationb_1,\belief),\state_i)\right]\right]+\transfer_{\joint\conjugate}(\state_{i-1}).
\end{align*}
Second, we construct an $N\times|\supp\;\bsplit\conjugate|$ matrix, $\beliefmatrix_{\bsplit\conjugate}$, with $(i,j)$-element $\bsplit\conjugate(\belief_j)\frac{\belief_j(\state_i)}{\prior(\state_i)}$.
%
%
%

Thus, finding transfers that implement the solution to the relaxed program reduces to verifying that a solution exists to $ \beliefmatrix_{\bsplit\conjugate} \tilde{\transfer}=\transfer_{\joint\conjugate}$, where $\tilde{\transfer}$ is a vector of transfers, one for each posterior in the support of \bsplit\conjugate.

\begin{prop}\label{prop:relaxed-more-conditions}
Suppose the solution to the relaxed program, $\joint\conjugate=(\bsplit\conjugate,\alpha\conjugate,\allocation_2\conjugate)$, satisfies the monotonicity constraints. Then, transfers $\tilde{\transfer}\conjugate$ exist  such that $(\bsplit\conjugate,\alpha\conjugate,\tilde{\transfer}\conjugate,\allocation_2\conjugate)$ solve \ref{eq:opt-limited-commitment} if and only if a solution $\tilde{\transfer}\in\reals^{|\supp\;\bsplit\conjugate|}$ exists to 
\[\beliefmatrix_{\bsplit\conjugate}\tilde{\transfer}=\transfer_{\joint\conjugate}.\]
That is, if and only if $\mathrm{rank}(\beliefmatrix_{\bsplit\conjugate})=\mathrm{rank}(\beliefmatrix_{\bsplit\conjugate}|\transfer_{\joint\conjugate})$.
\end{prop}
The conditions of \autoref{prop:relaxed-more-conditions} are satisfied, for instance, if the posterior distribution that solves the relaxed program induces $N$ linearly independent posteriors, or if it has singleton support. In the first case, the matrix $\beliefmatrix_{\bsplit\conjugate}$ is invertible, whereas in the second case, all types have the same allocation and pay the same transfer.

\begin{proof}[Proof of \autoref{prop:relaxed-more-conditions}] Let $\joint\conjugate=(\bsplit\conjugate,\tilde{\alpha}\conjugate,\allocation_2\conjugate)$ denote the solution to the relaxed program and let $M=|\supp\;\bsplit\conjugate|$. By \autoref{prop:vs}, $M\leq N$. Given the steps in  the proof of \autoref{prop:med}, it is immediate that if we can find transfers $\{\tilde{x}(\belief):\belief\in\supp\;\bsplit\conjugate\}$ that satisfy the constraints of the relaxed program, then we have a solution to \ref{eq:opt-limited-commitment}. 

Evidently, if $\tilde{\transfer}\in\reals^M$ exists such that $\beliefmatrix_{\bsplit\conjugate}\tilde{\transfer}=\transfer_{\joint\conjugate}$, then $(\bsplit\conjugate,\tilde{\alpha}\conjugate,\tilde{\transfer},\allocation_2\conjugate)$ solve \ref{eq:opt-limited-commitment}.
%
%
To get the second part of the if and only if, suppose that transfers $\tilde{\transfer}\in\reals^M$ exist such that  $(\bsplit\conjugate,\tilde{\alpha}\conjugate,\tilde{\transfer},\allocation_2\conjugate)$ solve \ref{eq:opt-limited-commitment}. Then, note that the following holds. First, for $\state_1$,
\begin{align*}
\mathbb{E}_{\bsplit\conjugate}\left[\frac{\belief(\state_1)}{\prior(\state_1)}\tilde{\transfer}(\belief)\right]=\mathbb{E}_{\bsplit\conjugate}\left[\mathbb{E}_{\tilde{\alpha}\conjugate}\left[\frac{\belief(\state_1)}{\prior(\state_1)}u(\allocationb_1,\allocation_2\conjugate(\allocationb_1,\belief),\state_1)\right]\right].
\end{align*}
Furthermore, for $i\geq2$
\begin{align*}
\mathbb{E}_{\bsplit\conjugate}\left[\frac{\belief(\state_i)}{\prior(\state_i)}\tilde{\transfer}(\belief)\right]=\mathbb{E}_{\bsplit\conjugate}\left[\mathbb{E}_{\tilde{\alpha}\conjugate}\left[\left(\frac{\belief(\state_i)}{\prior(\state_i)}-\frac{\belief(\state_{i-1})}{\prior(\state_{i-1})}\right)u(\allocationb_1,\allocation_2\conjugate(\allocationb_1,\belief),\state_i)\right]\right]+\mathbb{E}_{\bsplit\conjugate}\left[\frac{\belief(\state_{i-1})}{\prior(\state_{i-1})}\tilde{\transfer}(\belief)\right].
\end{align*}
This implies that $\tilde{\transfer}$ solves the system  $\beliefmatrix_{\bsplit\conjugate}\tilde{\transfer}=\transfer_{\joint\conjugate}$. 

The Rouché-Capelli theorem (Theorem 2.38 in \citealp{shafarevich2012linear}) then implies the rank conditions in the statement.
\end{proof}
\section{Proofs of \autoref{sec:informed-receiver}}\label{appendix:informed-receiver}
Without loss of generality, a menu of experiments consists of a finite set of signals $S$ and a collection of distributions $\{\pi_\type:\States\mapsto\Delta(S):\type\in\Types\}$. Under experiment $\pi_\type$, when the agent observes signal $s\in S$, the agent updates her belief about the state of the world as follows:
\begin{align*}
\belief_s(\state)=\frac{\prior(\state)\pi_\type(s|\state)}{\sum_{\stateb\in\States }\pi_\type(s|\state)\prior(\stateb)}\equiv\frac{\prior(\state)\pi_\type(s|\state)}{Pr_{\experiment,\type}(s)}.
\end{align*}
A menu of experiments is incentive compatible if the following holds for all $\type\in\Types$ and $\typeb\neq\type$:
\begin{align}\label{eq:ic-signals}
\sum_{\belief\in\Posteriors}\sum_{\{s\in S:\belief_s=\belief\}}Pr_{\experiment,\type}(s)U(\belief,\type)\geq\sum_{\belief\in\Posteriors}\sum_{\{s\in S:\belief_s=\belief\}}Pr_{\experiment,\typeb}(s)U(\belief,\type).
\end{align}
\begin{lemma}\label{lemma:rp}
It is without loss of generality to focus on experiments such that $S=\Posteriors$.
\end{lemma}
\begin{proof}
The statement follows from \autoref{eq:ic-signals}. To see this, let $\langle\{\pi_\type\}_{\type\in\Types},S\rangle$ denote an experiment. Consider the following experiment, $\langle\{\experimentb_\type\}_{\type\in\Types},\Posteriors\rangle$\begin{align}
\experimentb_\type(\belief|\state)=\sum_{\{s\in S:\belief_s=\belief\}}\pi_\type(s|\state).
\end{align}
Note that 
\begin{align*}
Pr_{\experimentb,\type}(\belief)=\sum_{\state\in\States}\prior(\state)\experimentb_\type(\belief|\type)=\sum_{\state\in\States}\prior(\state)\sum_{\{s\in S:\belief_s=\belief\}}\pi_\type(s|\state)=\sum_{\{s\in S:\belief_s=\belief\}}Pr_{\experiment,\type}(s).
\end{align*}
Thus, $\langle\{\experimentb_\type\}_{\type\in\Types},\Posteriors\rangle$ yields the same payoff to the designer and the agent. Furthermore, it is incentive compatible.
\end{proof}
\begin{proof}[Proof of \autoref{prop:informed-receiver}]
The proof proceeds in two steps. We first argue that the problems in \autoref{eq:opt-informed-receiver} and \ref{eq:optbis} have the same value. We then apply \autoref{theorem:tomalalemma} to the problem in \ref{eq:optbis} to argue for the upper bound in the number of posteriors induced in an optimal experiment.

To see that both problems have the same value\label{page-value}, consider the following argument. Let $\tau^*$ denote a solution to \autoref{eq:opt-informed-receiver}. For each $\type\in\Types$, let 
\begin{align*}
u_\type^*=\mathbb{E}_{\tau^*(\type,\cdot)}[U(\belief,\type)].
\end{align*}
Then, it is immediate to check that $(\tau^*,(u_\type^*)_{\type\in\Types})$ solves the problem in \ref{eq:optbis}.

Let $(\tau^*,(u_\type^*)_{\type\in\Types})$ denote a solution to the problem in \ref{eq:optbis}. Note that without loss of generality we can take 
\begin{align*}
u_\type^*=\mathbb{E}_{\tau^*(\type,\cdot)}[U(\belief,\type)].
\end{align*}
Note that for each \type\ this relaxes the incentive compatibility constraint for $\typeb\neq\type$ and it does not affect the first constraint for \type's experiment. It then follows that $\tau^*$ solves the problem in \autoref{eq:opt-informed-receiver}.

Consider now the problem in \ref{eq:optbis}. Fix $\{u_\type\}_{\type\in\Types}$. Note that the problem of finding an optimal $\tau: \Types \to \Delta_{\prior}\Posteriors$ given $\{u_\type\}_{\type\in\Types}$ is separable across $\type\in\Types$. That is, given $\{u_\type\}_{\type\in\Types}$, it is enough to solve $M$ optimization problems:
\begin{align}\label{eq:opt-type}
&\max_{\tau: \Types \to \Delta_{\prior}\Posteriors}\mathbb{E}_{\tau(\type,\cdot)}\left[V(\belief,\type)\right]\\
&\text{s.t.}\left\{\begin{array}{ll}&\mathbb{E}_{\tau(\type,\cdot)}\left[U(\cdot,\type)\right]\geq u_\type\\
(\forall\typeb\neq\type)&u_{\typeb}\geq\mathbb{E}_{\tau(\type,\cdot)}\left[U(\belief,\typeb)\right]
\end{array}\right.,\nonumber
\end{align}
where $V(\belief,\type)=\sum_{\state\in\States}\belief(\state)v(\action^*(\belief,\type),\type,\state)$. For each \type, the problem in \autoref{eq:opt-type} is a special case of the problem in \ref{eq:opt}. \autoref{cor:bound} implies that a solution exists that uses at most $N+M$ posteriors.
\end{proof}
\section{Omitted examples}\label{appendix:examples}
\subsection{The concave hull and the concave closure of \genericf\ may differ}\label{appendix:cav-not-clcav}
\begin{example}[\cav\;\genericf\ and \clcav\;\genericf\ may differ]\label{example:cav-not-clcav}
Consider the function
\begin{align}
\genericf(\variable)=\left\{\begin{array}{ll}0&\text{ if } \variable>0\\
-\infty&\text{otherwise}\end{array}\right..
\end{align}
Then, 
\begin{align}\tag{cav}
\cav(\genericf)(\variable)=\left\{\begin{array}{ll}0&\text{ if } \variable>0\\
-\infty&\text{otherwise}\end{array}\right.,
\end{align}
whereas
\begin{align}\tag{\clcav}
\clcav(\genericf)(\variable)&=\left\{\begin{array}{ll}0&\text{ if } \variable\geq0\\
-\infty&\text{otherwise}\end{array}\right..
\end{align}
\end{example}
\subsection{Refined upper bounds }\label{appendix:refined}
\autoref{example:sm} below illustrates how under the conditions of \autoref{prop:gid} one can obtain posterior distributions that induce more posteriors than the number of states $N$, but less than the number of states plus constraints, $N+\inequality+\equality$. In the example, $N=\inequality=2$ and both constraints bind at the optimum. However, (\objective,\vectorconstraint) is a 3-generalized information design environment.
\begin{example}[Social Media]\label{example:sm} A social media platform designs how information is released to consumers and wishes to be perceived as unbiased. We represent this as follows. There are two equally likely states of the world, $\States=\{\state_L,\state_R\}$, where $\state_i$ denotes the platform's political inclination. The platform's payoff is given by
\begin{align}\label{eq:obj-social-media}
\objective(\belief)=\frac{1}{2}-\left|\frac{1}{2}-\belief\right|,
\end{align}
where $\belief\in[0,1]$ is the likelihood an outside observer attaches to the platform being right-leaning $\state=\state_R$ after spending time on it. Consistent with its desire to be perceived as neutral, the platform's payoffs are maximized when consumers do not learn anything relative to the prior about its political inclinations.

Similar to \autoref{example:news}, the platform must collect ad revenue to operate and for this it requires a broad audience. A left (right) leaning consumer enjoys spending time on the platform only if she perceives the platform's content as left (right) leaning, which we model by the following payoff functions:
\begin{align}\label{eq:const-social-media}
\const_L(\belief)=\max\left\{\frac{1}{4}-\belief,0\right\}, &\const_R(\belief)=\max\left\{0,\belief-\frac{3}{4}\right\}.
\end{align}
A left (right) leaning consumer participates on the platform if her expected payoff is larger than $\side_L$ ($\side_R$). \autoref{fig:sm-obj-cons} depicts the platform's and the audiences' payoff functions.
\begin{figure}[t!]
\centering
\subfloat[Objective function]{\begin{tikzpicture}[scale=0.8]
\begin{axis}[axis lines=middle,xmin=-0.05,xmax=1.1,ymin=-0.05,ymax=0.75,xtick={0,0.25,0.5,0.75,1},xticklabels={0,$\frac{1}{4}$,$\prior$,$\frac{3}{4}$,1},ytick={0,0.5,1},xlabel=$\belief$,ylabel=$\objective$,   x label style={at={(axis description cs:1,-0.05)}},
    y label style={at={(axis description cs:-0.05,1)}},
    width=8cm,
        height=8cm]
\addplot[color=black,thick]coordinates {(0,0) (0.5,0.5)};
\addplot[color=black,thick]coordinates {(1,0) (0.5,0.5)};
\end{axis}\end{tikzpicture}}
\subfloat[Constraints]{\begin{tikzpicture}[scale=0.8]
\begin{axis}[axis lines=middle,xmin=-0.05,xmax=1.1,ymin=-0.05,ymax=0.75,xtick={0,0.25,0.5,0.75,1},xticklabels={0,$\frac{1}{4}$,$\prior$,$\frac{3}{4}$,1},ytick={0,0.5,1},xlabel=$\belief$,ylabel=$\const_\cdot$,   x label style={at={(axis description cs:1,-0.05)}},
    y label style={at={(axis description cs:-0.05,1)}},
    width=8cm,
        height=8cm,legend style={fill=none}]
\addplot[color=blue,thick]coordinates {(1,0) (0.25,0)};
\addlegendentry{$\const_L$};
\addplot[color=red,thick,domain=0.75:1]{x-0.75};
\addlegendentry{$\const_R$};
\addplot[color=red,thick]coordinates {(0,0) (0.75,0)};
\addplot[color=blue,thick]coordinates {(0,0.25) (0.25,0)};
\end{axis}\end{tikzpicture}}
\caption{Objective function (left) and constraints (right) in \autoref{example:sm}}\label{fig:sm-obj-cons}
\end{figure}
Without participation, the platform gets no revenues so the optimal information disclosure policy solves:
\begin{align}\tag{\OPT$_{SM}$}\label{eq:opt-social-media}
\max_{\bsplit\in\Delta_{\prior}\Posteriors}&\mathbb{E}_\bsplit\left[\objective\right]\\
\text{ s.t. }&\mathbb{E}_\bsplit\left[\const_i\right]\geq\side_i,i\in\{L,R\}.\nonumber
\end{align}
Note that the tuple $(\objective,\const_L,\const_R)$ is a 3-generalized information design environment with partition $\Delta_1=[0,\nicefrac{1}{4}),\Delta_2=[\nicefrac{1}{4},\nicefrac{3}{4}),\Delta_3=[\nicefrac{3}{4},1]$.

%

\autoref{prop:lagrangian} implies that in order to solve the platform's problem we can consider the following Lagrangian objective function:
\begin{align}\label{eq:lagrangian-social-media}
\left(\objective+\dual\const\right)(\belief)&=\left\{\begin{array}{ll}\belief+\dual_L\left(\belief-\frac{1}{4}\right)&\text{ if }\belief\leq0.25\\
\belief&\text{ if }\belief\in(0.25,0.5]\\
1-\belief&\text{ if }\belief\in(0.5,0.75)\\1-\belief+\dual_R\left(\belief-\frac{3}{4}\right)&\text{ otherwise }\end{array}\right..
\end{align}
It is possible to show that the optimal solution corresponds to $\dual_L\conjugate=\dual_R\conjugate=2$ and the platform's content policy having support in $\{0,\nicefrac{1}{2},1\}$. That is, the platform balances its desire to appear neutral, with enough political content to attract its audience. This can be confirmed by visually inspecting the Lagrangian objective function and its concavification at the prior in \autoref{fig:sm-lagrangian} for different values of the multipliers $\dual_L,\dual_R$. 


\begin{figure}[t!]
\centering
\subfloat[$\dual_L=\dual_R=1<\dual\conjugate$]{
\begin{tikzpicture}[scale=0.75]
\begin{axis}[axis lines=middle,xmin=-0.05,xmax=1.1,ymin=-0.05,ymax=0.8,xtick={0,0.25,0.5,0.75,1},xticklabels={0,$\frac{1}{4}$,$\prior$,$\frac{3}{4}$,1},ytick={0,0.5,1},xlabel=$\belief$,ylabel=$\objective+\dual\const$,   x label style={at={(axis description cs:1,-0.05)}},
    y label style={at={(axis description cs:-0.15,1)}},
    width=8cm,
        height=8cm]
\addplot[thick,domain=0:0.25]{x+1*(0.25-x)};
\addplot[thick,domain=0.25:0.5]{x};
\addplot[thick,domain=0.5:0.75]{1-x};
\addplot[thick,domain=0.75:1]{1-x+1*(x-0.75)};
\addplot[color=red,dashed,thick] coordinates {(0,0.25) (0.5,0.5)};
\addplot[color=red,dashed,thick] coordinates {(1,0.25) (0.5,0.5)};
\end{axis}
\end{tikzpicture}\label{fig:sm-low-dual}}
\subfloat[$\dual_L=\dual_R=\dual\conjugate=2$]{\begin{tikzpicture}[scale=0.75]
\begin{axis}[axis lines=middle,xmin=-0.05,xmax=1.1,ymin=-0.05,ymax=0.8,xtick={0,0.25,0.5,0.75,1},xticklabels={0,$\frac{1}{4}$,$\prior$,$\frac{3}{4}$,1},ytick={0,0.5,1},xlabel=$\belief$,ylabel=$\objective+\dual\conjugate\const$,   x label style={at={(axis description cs:1,-0.05)}},
    y label style={at={(axis description cs:-0.18,1)}},
    width=8cm,
        height=8cm]
\addplot[thick,domain=0:0.25]{x+2*(0.25-x)};
\addplot[thick,domain=0.25:0.5]{x};
\addplot[thick,domain=0.5:0.75]{1-x};
\addplot[thick,domain=0.75:1]{1-x+2*(x-0.75)};
\addplot[color=red,dashed,thick] coordinates {(0,0.5) (1,0.5)};
\end{axis}
\end{tikzpicture}\label{fig:sm-opt-dual}}
\subfloat[$\dual_L=\dual_R=3>\dual\conjugate$]{\begin{tikzpicture}[scale=0.75]
\begin{axis}[axis lines=middle,xmin=-0.05,xmax=1.1,ymin=-0.05,ymax=0.8,xtick={0,0.25,0.5,0.75,1},xticklabels={0,$\frac{1}{4}$,$\prior$,$\frac{3}{4}$,1},ytick={0,0.5,1},xlabel=$\belief$,ylabel=$\objective+\dual\const$,   x label style={at={(axis description cs:1,-0.05)}},
    y label style={at={(axis description cs:-0.15,1)}},
    width=8cm,
        height=8cm]
\addplot[thick,domain=0:0.25]{x+3*(0.25-x)};
\addplot[thick,domain=0.25:0.5]{x};
\addplot[thick,domain=0.5:0.75]{1-x};
\addplot[thick,domain=0.75:1]{1-x+3*(x-0.75)};
\addplot[color=red,dashed,thick] coordinates {(0,0.75) (1,0.75)};
\end{axis}
\end{tikzpicture}\label{fig:sm-high-dual}}
\caption{Lagrangian approach in \autoref{example:sm}; \dual\conjugate\ denotes the optimal multiplier. The dashed red line is the concavification.}\label{fig:sm-lagrangian}
\end{figure}
\end{example}

\end{document}